\documentstyle[epsf,aps,prb,eqsecnum,preprint]{revtex}

\begin{document}

\title{Charge gap in the one--dimensional dimerized Hubbard
model at quarter-filling}
\author{Karlo Penc\cite{szfki}}
\address{
 {\it Institut de Physique, Universit\'e de Neuch\^atel }\\
 {\it Rue Breguet 1, CH-2000 Neuch\^atel, Switzerland  }
}
\author {Fr\'ed\'eric Mila}
\address{
  Laboratoire de Physique Quantique\\
   Universit\'e Paul Sabatier, 31062 Toulouse, France
}

\date{\today}

\maketitle
\begin{abstract}
  We propose a quantitative estimate of the charge gap that opens in the
one-dimensional dimerized Hubbard model at quarter-filling
due to dimerization, which makes the
system effectively half--filled, and to repulsion, which induces umklapp
scattering processes. Our estimate is expected to be valid for any value of the
repulsion and of the parameter describing the dimerization.
It is based on analytical results obtained in various limits (weak coupling,
strong coupling, large dimerization) and on numerical results obtained by exact
diagonalization of small clusters. We consider two models of dimerization:
alternating hopping integrals and alternating on--site energies. The former
should be appropriate for the Bechgaard salts, the latter for compounds
where the stacks are made of alternating $TMTSF$ and $TMTTF$ molecules.
\end{abstract}


\section{Introduction}

A large number of compounds exhibit one--dimensional electronic properties that
can
be described by a quarter--filled Hubbard model with some kind of dimerization.
For instance,  the Bechgaard salts $(TMTSF)_2 X$ and $(TMTTF)_2 X$, where $X$
denotes
  an anion like ${ClO_4}^-$, ${PF_6}^-$, ${Br}^-$, etc... can be regarded
concerning their
electronic structure as being essentially one--dimensional systems above a
crossover temperature $T_x$ of the order of 30K\cite{jerome}.
{}From stoichiometry it is known that there are 3 electrons in the HOMO for
each
pair $(TMTSF)_2$, so that the system is 3/4--filled in terms of electrons or
quarter--filled in terms of holes. In the following we will always use hole
notation and consider quarter--filled systems. A reasonable description of
these properties should be provided by the dimerized Hubbard model
 defined by the Hamiltonian
\begin{equation}
H=H_0+H_{\rm int}
  \label{eq:Ham}
 \;,
\end{equation}
where the kinetic part of the Hamiltonian is (see
Fig.~\ref{fig:model})
\begin{equation}
 H_0= -t_1 \sum_{i {\rm even}, \sigma} (c_{i,\sigma}^\dagger
      c_{i+1,\sigma}^{\vphantom{\dagger}} + {\rm h.c.})
      -t_2 \sum_{i {\rm odd}, \sigma} (c_{i,\sigma}^\dagger
       c_{i+1,\sigma}^{\vphantom{\dagger}} + {\rm h.c.})
   \label{eq:Ham_t_alter}
 \;.
\end{equation}
 The
operators $c_{i,\sigma}^\dagger$ create particles in the HOMO of $TMTSF$ or
$TMTTF$
with spin $\sigma$. We have included two hopping integrals $t_1$ and $t_2$
($t_1>t_2$) to describe the dimerization along the
stacks\cite{gallois,ducasse}.
The dispersion of
this model is given by $\varepsilon(k)=\sqrt{t_1^2+t_2^2+2 t_1 t_2 \cos k}$ and
is depicted in Fig.~\ref{fig:bands}. The important parameters are the
dimerization
gap $\Delta_D=2(t_1-t_2)$ that opens at the Brillouin zone boundary and the
total bandwidth
$W=2(t_1+t_2)$.

To the hopping part of the Hamiltonian we have to add an interaction part to
describe the correlation between the electrons. For simplicity, we have chosen
the form of the standard Hubbard model
\begin{equation}
  H_{\rm int}= U \sum_{i} n_{i\uparrow}n_{i\downarrow}
  \label{eq:Ham_int}
 \;,
\end{equation}
 where $U$ is the on-site Coulomb repulsion and
$n_{i,\sigma}=c_{i,\sigma}^\dagger c_{i,\sigma}^{\vphantom{\dagger}}$.

This model should also provide a good description of salts like $(FA)_2PF_6$,
where $FA$ stands for Fluoranthenyl, which undergo a Peierls transition
leading to dimerization along the stacks at a temperature of order
190K\cite{krohnke,sachs}.
There is another class of materials however for which a modified version of the
previous model is more appropriate. These materials are related to the
Bechgaard salts, but the stacks now consist of alternating
$TMTTF$ and $TMTSF$ molecules\cite{pouget}. A minimal model in this case would
be a Hubbard
model with alternating atomic on--site energies, with a kinetic part of the
form
\begin{equation}
H_0 = -t \sum_{i , \sigma} (c_{i,\sigma}^\dagger
      c_{i+1,\sigma}^{\vphantom{\dagger}} + {\rm h.c.})
    -\varepsilon_0 \sum_{i {\rm even}, \sigma} (c_{i,\sigma}^\dagger
      c_{i,\sigma}^{\vphantom{\dagger}}
      - c_{i+1,\sigma}^\dagger
       c_{i+1,\sigma}^{\vphantom{\dagger}})
   \label{eq:Ham_e_alter}
 \;,
\end{equation}
where $2\varepsilon_0$ is the energy splitting between the HOMO of the $TMTSF$
and $TMTTF$. The dispersion relation is given by
$\varepsilon(k)=\sqrt{\varepsilon_0^2+2 t^2 + 2 t^2 \cos k}$. It has
the same form as in
the previous case. The dimerization gap is now given
by $\Delta_D=\sqrt{\varepsilon_0^2+4 t^2}$ and the bandwidth by
$W=2\varepsilon_0$. However, although the dispersion relations are the same,
the two models are  different because the interaction part will not be  the
same when expressed in terms of the operators that diagonalize the kinetic part
of the Hamiltonian.
 Let us also note that a similar model has been
recently proposed by
Sudb\o~{\it et al.}\cite{sudbo}
as a one dimensional analog of the copper oxide layers in the high temperature
superconductors. A model where
dimerization is induced via alternating on--site repulsions
has also been studied\cite{Japaridze}.

   It is clear from Fig.~\ref{fig:bands} that, due to dimerization, the
system is effectively half--filled. In this case, it is known from the general
theory of one--dimensional models\cite{Solyom,Emery} with umklapp scattering
that an on--site
repulsion of arbitrary size will open a gap in the charge sector. This has been
shown explicitly in the case of the Hubbard model for which exact results are
available from the Bethe ansatz solution\cite{LiebWu}. In the general case,
where no exact
solution is known, a quantitative estimate of the gap is not available so
far.
This is unfortunate because such an estimate is necessary to interprete
the activated behaviour of the resistivity observed at relatively low
temperatures
in several compounds, which in turn provides an estimate of the magnitude
of the Coulomb repulsions. An analysis of that sort has already been
performed for the Bechgaard salts of the $TMTTF$ family
on the basis of a preliminary determination of the charge gap in
the model with alternating hopping amplitudes\cite{penc}.

In this paper, we propose a quantitative estimate of the gap in the case
of
the dimerized models described above that should be valid in the whole range
of parameters.
 This estimate is based on approximations that provide analytical expressions
in various limits, and on numerical calculations using L\'anczos
diagonalization
of small clusters. As we think that the results might
be useful to readers that do not want to go through the details of the
calculation, we start in Sec.~II with a detailed account of the results,
including
analytical expressions, tables and curves, that can be used independently of
the rest of the paper. Then, in Sec.~III we explain how and for which values of
 the parameters numerical estimates
can be obtained on the basis of exact diagonalization of small clusters. The
following sections are devoted to the approximate expressions that can be
obtained analytically in various limits: Sec.~IV summarizes exact results in a
few trivial limits, Sec.~V deals with the large dimerization case, where the
dimerization gap is larger than both the repulsion and the Fermi velocity, and
Secs.~VI and VII with the case of weak and strong coupling, respectively.
The summary can be found in Section~II. Throughout the paper,
subsections A will deal with the model with alternating hopping amplitudes, and
subsections B with the model with alternating on--site energies

\section{Summary of results}

  In Fig.~\ref{fig:region} we have illustrated the different regimes
in which analytical results have been obtained
as a
function of the model parameters.
In this section, we give a summary of these results for each model.

\subsection{Alternating hopping amplitudes}

  \subsubsection{Numerical estimates:}
    Using exact diagonalization of small clusters, we have obtained estimates
of the charge gap for intermediate values of repulsion $U$ and for
$t_2 \leq 0.5 t_1$. The limitation comes from the size of the clusters
(maximum 16
sites), so that only large gaps corresponding to correlation lengths smaller
that 16 sites could be extracted.
  The numerical results are shown in Fig.~\ref{fig:appr}a. They
are in good agreement with analytical approximations in different
limits, as we will show in the subsequent paragraphs.

 \subsubsection{Large dimerization:}
   In this limit we have mapped our model onto the exactly solvable Hubbard
model
at half--filling and we found that the gap is given by:
\begin{equation}
 \Delta_c= \left\{
    \begin{array}{ll}
      \displaystyle{{t_2 \over 2} {8\over \pi} \sqrt{U \over t_2}
       e^{-2\pi t_2/U }
       e^{-\pi t_2/4t_1 }} & \mbox{, if $t_1 \gg t_2 \gg U$;} \\
       &\\
      \displaystyle{{U\over 2} - {U^2\over 16t_1}-2t_2}
       & \mbox{, if $t_1 \gg U \gg t_2$.}
    \end{array}
   \right.
  \label{eq:gapt2st}
\end{equation}
 In Fig.~\ref{fig:largedim} we have compared our estimate with numerical
results (curve $a$) for $t_2=0.1$, where we can see the crossover from the
exponential regime (curve $d$) to the linear one
(curves $b$ and $c$, $b$ is without the quadratic term  $U^2/16t_1$) for
$U/t_1 \sim 0.5$ .

\subsubsection{Weak coupling limit}
In the weak coupling limit ($U\ll t_1,t_2$), using the RG group
method and results from the large dimerization limit, we found that the
gap is exponentially small [see Eq.~(\ref{eq:gap_sUArd})]:
\begin{equation}
   \Delta_c = a t_1 \sqrt{U\over t_1} \exp\left(-{b t_1\over U}\right)
  \label{eq:gap_sUA}
 \;,
\end{equation}
where the parameters $a$ and $b$ can be found in Table~\ref{tab:althop} for
different values of $t_2/t_1$.

  In Fig.~\ref{fig:gapweak} we compare the numerical values of the gap for
$t_2=0.3$ and $t_2=0.4$  with the weak coupling approximation
Eq.~(\ref{eq:gap_sUA}). We can see that up to $U/t_1=4$ the weak coupling
formula
gives very good results.

 \subsubsection{Strong coupling limit}

When the on--site repulsion is large enough ($U \gg t_1, t_2, 4 t_1^2/t_2$),
the
effective Hamiltonian of our model is a $t-J$ like Hamiltonian. Applying
degenerate perturbation theory, we found that the charge gap is equal to the
dimerization gap with a $1/U$ correction [see Eq.~(\ref{eq:gap_lUAnd})]:
\begin{equation}
  \Delta_c = \Delta_D \left(1- c{ t_1 \over U}\right)
  \label{eq:gap_lUA}
 \;,
\end{equation}
where the parameters $\Delta_D/t_1$ and $c$ can be found for different values
of $t_2/t_1$ in Table~\ref{tab:althop}.

\subsubsection{Intermediate region}

On the basis of the previous results, we can propose an estimate of the gap
for any value of the parameters. This can be achieved by using Pad\'e
approximants to connect the exact results we have obtained for $U$ small and
$U$ large respectively.
  For $t_2\geq 0.2$, we have used the weak coupling expression up to $U_0$,
where $U_0$
is not too large (actually we choose $U_0/t_1=b$ for $t_2/t_1\leq 0.5$ and
$U_0/t_1=4$ for $t_2/t_1> 0.5$) and Pad\'e approximants of the form
\begin{equation}
   f(U) = { a_1 U + \Delta_D U^2 \over b_2 + b_1 U + U^2}
   \label{eq:pade}
\end{equation}
for  $U>U_0$. The coefficients are chosen in such a way that, for
$U\rightarrow + \infty$, the Pad\'e approximant gives the correct large $U$
behavior given by Eq.~(\ref{eq:gap_lUA}) and that, at $U=U_0$, the resulting
curve
and its derivative be continuous. The curves depend weakly on the value of
$U_0$: Changing the value of $U_0$ by 50\% affects the value of the gap by
less then 10\%.  In Table~\ref{tab:pade} we give the values of the coefficients
and of $U_0$ for different values of $t_2/t_1$.

  For $t_2 = 0.1 t_1$ we have also used a Pad\'e approximant, but now the
limiting behaviors were determined by the second equation in
(\ref{eq:gapt2st}),
which is valid for the intermediate values of $U$ for $t_2$ small, and by
Eq.~(\ref{eq:gap_lUA}) in the large $U$ limit.
At the crossing point of the weak-coupling expression and the Pad\'e
approximant, we have got rid of the kink (see Fig.~\ref{fig:largedim}) by a
linear interpolation.
 The resulting
curves for $t_2/t_1$ ranging from 0.1 to 0.9 are presented in
Fig.~\ref{fig:appr}.

\subsection{Alternating on--site energies}

 \subsubsection{Large dimerization:}
   In this limit we have mapped our model to the exactly solvable Hubbard model
at half--filling and we found that the gap is given by:
\begin{equation}
 \Delta_c= \left\{
    \begin{array}{ll}
      \displaystyle{
	{8\over \pi} t \sqrt{U \over 2 \varepsilon_0}
	 \exp(-\pi t^2/U \varepsilon_0)}
       & \mbox{, if $\epsilon_0 \gg t \gg U$;} \\
       &\\
       U - \displaystyle{2t^2 \over \varepsilon_0}
       & \mbox{, if $\epsilon_0 \gg U \gg t$.}
    \end{array}
   \right.
\end{equation}

\subsubsection{Weak coupling limit}
In the weak coupling limit ($U\ll \varepsilon_0,t$), on the basis of
renormalization group
analysis and results from the large dimerization limit, the
gap is exponentially small and is
given by [see also Eq.~(\ref{eq:gap_smallUB})]:
\begin{equation}
  \Delta_c  = a t \sqrt{U\over t} \exp\left(-b {t\over U} \right)
 \;.
\end{equation}
 The parameters $a$ and $b$ are given in Table~\ref{tab:altene} for some values
of $\varepsilon_0/t$.

 \subsubsection{Strong coupling limit}

When the on--site repulsion is large enough ($U \gg \epsilon_0, t,
4t^2/\epsilon_0$), the
effective Hamiltonian of our model is again a $t-J$ like Hamiltonian and
 we found that the charge gap is [see also Eq~(\ref{eq:gaplargeUB})]:
\begin{equation}
    \Delta_c = \Delta_D \left(1- c{ t \over U} \right)
 \;.
\end{equation}
 The parameters $\Delta_D$ and $c$ are given in Table~\ref{tab:altene} for some
values of $\varepsilon_0/t$.

\section{Numerical Simulations}

  We have done intensive numerical simulations based on L\'anczos
diagonalization of small clusters for the model with alternating hopping
integrals. Our numerical estimation for the gap was based on the usual formula
\begin{eqnarray}
  \Delta_c &=& \lim_{N \rightarrow +\infty} \Delta_c(N) \;, \nonumber\\
  \Delta_c(N) &=& E(N+1;2N)+E(N-1;2N)-2E(N;2N) \;,
 \label{eq:numgap}
\end{eqnarray}
where $E(M;L)$ is the ground state energy for $M$ particles on $L$ sites. We
have obtained results for systems of 4, 8, 12 and 16 sites.
Periodic (resp. antiperiodic) boundary conditions have been used for L=8 and
16 (resp. L=4 and 12) to have open-shell systems.
Typical results are
shown on Fig.~\ref{fig:gapsca}, where we have plotted $\Delta_c(N)$ as a
function of $1/N$. Good estimates can be obtained only when the exponential
regime $\Delta_c(N)-\Delta_c \sim \exp(- {\rm cst}/N)$ has been reached for
$N=16$. But the exponential regime is obtained when $v_c/N < \Delta_c$, that is
$N> v_c/\Delta_c$. So we can only obtain good numerical estimates of the gap
when $\Delta_c$ is not too small, i.e. when $U$ and $\Delta_D$ are not too
small.
The gap can then be extracted by use of the Shanks transformation\cite{bender}
given by
\begin{equation}
\Delta_c={\Delta(N_1)\Delta(N_3)-\Delta(N_2)^2 \over \Delta(N_1)+\Delta(N_3)-
2\Delta(N_2)}
 \;.
\end{equation}
This procedure has been shown to work very well in the case of the Haldane
gap
of spin-1 chains\cite{golinelli}.
 Applying it for $(N_1,N_2,N_3)=(4,8,12)$ and $(8,12,16)$ respectively gave
similar
results. The results quoted throughout this paper are those obtained for
$(8,12,16)$ because they are a priori better.

  Another limitation comes from the size of the repulsion. When $U$ is very
large, the convergence of the L\'anczos algorithm becomes very slow, and it is
no longer possible to get good values of the energy. So, in summary, good
estimates have been obtained for $\Delta_D \gtrsim t_1$ and
$1 \lesssim U/t_1 \lesssim 60$. This does not cover the range of parameters for
actual compounds, for which one often has $\Delta_D/t_1$, or $U/t_1 \ll 1$, and
analytical methods are clearly needed to complement these numerical results. In
fact, the analytical methods developed in the rest of the paper provide an
estimate of the gap for any value of the parameters, and the numerical
simulations have been only used to check the analytical results.

\section{Exact results}
A schematic picture of the regions of interest as a function of the model
parameters are given in Fig.~\ref{fig:region} for each  model.  Exact, although
trivial, results can be deduced on the boundaries  and they are summarized in
this section.
\subsection{Alternating hopping amplitudes}

  (i) $t_1=t_2$: This is a quarter--filled Hubbard model and we know from the
Bethe ansatz solution that there is no gap in
the spectrum.

  (ii) $t_2=0$: The Hamiltonian describes a set of independent systems, each
system consisting of two sites. At 1/4 filling there is one electron for each
pair of sites in the ground state. The gap is then given by $\Delta_c=
E_0(0)+E_0(2)-2E_0(1)$, where $E_0(n)$ is the ground state energy of $n$
electron on a pair of sites. These energies are
$E_0(0)=0$, $E_0(1)=-2t_1$ and $E_0(2)=(U-\sqrt{U^2+16t_1^2})/2$
and the charge gap is
\begin{equation}
    \Delta_c= 2t_1+{U \over 2}-\sqrt{{U^2\over 4}+4 t_1^2} =
    \left\{
      \begin{array}{ll}
	 \displaystyle{{U \over 2} - {U^2\over 16 t_1}}&
	   \mbox{, if $t_1\gg U$;} \\
	 & \\
	 \displaystyle{2t_1-{4 t_1^2\over U}}& \mbox{, if $t_1\ll U$.}
      \end{array}
    \right.
      \label{eq:gapt2is0}
\end{equation}

  (iii) $U=0$: The band structure is given on Fig.~\ref{fig:bands}. The Fermi
energy is in the middle of the lower band, and $\Delta_c=0$.

  (iv) $U=+\infty$: The energies that enter Eq.~(\ref{eq:numgap}) are the same
as for free spinless fermions at half--filling, because the energy is
independent of the spin when $U=+\infty$ for open boundary conditions (or for
periodic one in the limit $N \rightarrow +\infty$). So, the charge gap is
actually the same as for a system of spinless fermions described by the kinetic
part of the Hamiltonian at half--filling, which is nothing but the dimerization
gap (see Fig.~\ref{fig:bands}). So $\Delta_c=\Delta_D$.

\subsection{Alternating on--site energies}

  (i) $\varepsilon_0=0$: This is a quarter filled Hubbard model, and there is
no gap.

  (ii) $\varepsilon_0=+\infty$: The odd sites are outside the Hilbert space.
The ground state has one particle per even site, and the first excited state
has a doubly occupied site. So clearly $ \Delta_c= U $.

  (iii) $U=0$: The Fermi energy is in the middle of the band, and $\Delta_c=0$.

  (iv) $U=+\infty$: The dimerization gap $2 \varepsilon_0$ and the charge gap
coincides, like in the case of alternating hoppings.

  (v) In the atomic limit ($t=0$) the charge gap is given by
$\Delta_c=\min(U,2\varepsilon_0)$.

\section{The limit of large dimerization}

  In this section, we consider the limit of large dimerization, that is the
limit where the
dimerization gap $\Delta_D$ is much larger than both the width of each subband
and the repulsion $U$. This limit corresponds to $t_1 \gg t_2,U$ for the model
with alternating hoppings and to $\varepsilon_0\gg t,U$ for the model with
alternating on--site energies. It is particularly interesting because in both
cases the model can be mapped onto the half--filled Hubbard model, so that we
can use the exact result provided by the Bethe Ansatz for the charge
gap\cite{LiebWu}.

  \subsection{Alternating hopping amplitudes}

  To diagonalize the part of the Hamiltonian corresponding to the hopping
terms $t_1$, we
introduce bonding and antibonding operators defined by:
\begin{eqnarray}
  b_{j,\sigma}^{\vphantom{\dagger}}&=&{1\over \sqrt{2}}
    \left[
      c_{j,\sigma}^{\vphantom{\dagger}} +
      c_{j+1,\sigma}^{\vphantom{\dagger}}
    \right] \;,\nonumber\\
  a_{j,\sigma}^{\vphantom{\dagger}}&=&{1\over \sqrt{2}}
    \left[
      c_{j,\sigma}^{\vphantom{\dagger}} -
      c_{j+1,\sigma}^{\vphantom{\dagger}}
    \right]
 \;,
\end{eqnarray}
where $j$ is now even. In terms of these operators,
the Hamiltonian becomes
\begin{eqnarray}
H = -t_1 \sum_{j {\rm even}, \sigma}&& \left(
   b_{j,\sigma}^\dagger b_{j,\sigma}^{\vphantom{\dagger}} -
   a_{j,\sigma}^\dagger a_{j,\sigma}^{\vphantom{\dagger}}
   \right) \nonumber\\
   -{t_2 \over 2} \sum_{j {\rm even}, \sigma}&& \left(
   b_{j,\sigma}^\dagger b_{j+2,\sigma}^{\vphantom{\dagger}} +
   b_{j+2,\sigma}^\dagger b_{j,\sigma}^{\vphantom{\dagger}} +
   b_{j,\sigma}^\dagger a_{j+2,\sigma}^{\vphantom{\dagger}} +
   a_{j+2,\sigma}^\dagger b_{j,\sigma}^{\vphantom{\dagger}} \right. \nonumber\\
   -&&\left.
   a_{j,\sigma}^\dagger b_{j+2,\sigma}^{\vphantom{\dagger}} -
   b_{j+2,\sigma}^\dagger a_{j,\sigma}^{\vphantom{\dagger}} -
   a_{j,\sigma}^\dagger a_{j+2,\sigma}^{\vphantom{\dagger}} -
   a_{j+2,\sigma}^\dagger a_{j,\sigma}^{\vphantom{\dagger}}
   \right)\nonumber\\
   + {U\over 2} \sum_{j{\rm even}} &&\left[
     \left(n_{d,j\uparrow}+n_{f,j\uparrow}\right)
     \left(n_{d,j\downarrow}+n_{f,j\downarrow}\right)+
     \left(
       b_{j,\uparrow}^\dagger a_{j,\uparrow}^{\vphantom{\dagger}} +
       a_{j,\uparrow}^\dagger b_{j,\uparrow}^{\vphantom{\dagger}}
     \right)
     \left(
       b_{j,\downarrow}^\dagger a_{j,\downarrow}^{\vphantom{\dagger}} +
       a_{j,\downarrow}^\dagger b_{j,\downarrow}^{\vphantom{\dagger}}
     \right)
      \right]
 \;.
\end{eqnarray}
The bonding and the antibonding bands are separated by a large energy $2t_1$.
The occupation of the upper band is thus negligible and to zeroth order in
$1/t_1$ we get
\begin{equation}
 H_{\rm eff} =
   -{t_2 \over 2}\sum_{j{\rm even},\sigma}
  (b_{j,\sigma}^\dagger b_{j+2,\sigma}^{\vphantom{\dagger}} + {\rm h.c.})
  +{U\over 2} \sum_{j{\rm even}} n_{b,j,\uparrow}n_{b,j,\downarrow}
   + O(1/t_1)
 \;,
\end{equation}
which is nothing but the regular Hubbard model with a repulsion $\tilde U=U/2$
and a hopping integral $\tilde t=t_2/2$.
 Here the Hamiltonin acts on a Hilbert space where the antibonding states are
all empty. The value of the gap in the
half--filled Hubbard model is known exactly from the Bethe Ansatz solution.
For small interaction ($\tilde U\ll \tilde t$), it is given by $\Delta_c=(8
\tilde t / \pi) \sqrt{\tilde U/\tilde t} \exp(-2\pi \tilde U/\tilde t)$, which,
with our notations, reads
\begin{equation}
    \Delta_c=
	 {t_2 \over 2} {8\over \pi}
      \sqrt{U \over t_2} e^{-2\pi t_2/U }
  \label{eq:gapt2small}
 \;.
\end{equation}
For large interaction
($U \gg t_2$ but still $U\ll t_1$), the exact expression becomes
$ \Delta_c= U/ 2 -2t_2$.

  To compare with other limits, it is useful to go to next order in $1/t_1$.
Using a Schrieffer--Wolff transformation, we can determine the $1/t_1$
corrections to the zeroth order effective Hamiltonian by including scattering
processes to the antibonding band and we get
\begin{eqnarray}
  H_{\rm eff} = &&
  -{t_2 \over 2}\sum_{j{\rm even},\sigma}
  (b_{j,\sigma}^\dagger b_{j+2,\sigma}^{\vphantom{\dagger}} + {\rm h.c.})
  +\left( {U\over 2}-{U^2\over 16 t_1} \right)
  \sum_{j{\rm even}} n_{b,j,\uparrow}n_{b,j,\downarrow}
  \nonumber\\
  &+&{t_2^2 \over 8 t_1}\sum_{j{\rm even},\sigma}
   \left(
     b_{j-2,\sigma}^\dagger b_{j+2,\sigma}^{\vphantom{\dagger}}
    +b_{j+2,\sigma}^\dagger b_{j-2,\sigma}^{\vphantom{\dagger}} \right)
   + O(1/t_1^2)
  \label{eq:effHubA2}
 \;.
\end{eqnarray}
The on--site repulsion is reduced by a $U^2/16 t_1$, and a second
nearest neighbor hopping appears. The formula for the gap is now modified to
\begin{equation}
 \Delta_c= \left\{
    \begin{array}{ll}
      \displaystyle{
	 {t_2 \over 2} {8\over \pi} \sqrt{U \over t_2}
	  e^{-2\pi t_2/U } e^{-\pi t_2/4t_1 }
      }
      & \mbox{, if $U \ll t_2$;} \\
      & \\
      \displaystyle{{U\over 2} - {U^2\over 16t_1}-2t_2}
      & \mbox{, if $U \gg t_2$.}
    \end{array}
   \right.
  \label{eq:gapt2}
\end{equation}
 The second formula agrees with Eq.~(\ref{eq:gapt2is0}) when $t_2=0$.
The first one
will be used in Section VI.
 We have illustrated the above estimates of the gap on Fig.~\ref{fig:largedim}.

\subsection{Alternating on--site energies}

  In the limit $\varepsilon_0 \gg U, t$ the occupation of the energetically
lower lying even sites is much larger than that of the odd sites. Using this,
we can again find an effective Hamiltonian starting from $t=0$, in which case
 only the
even sites are occupied. Switching on the hopping, the electrons can hop to the
energetically unfavorable odd sites, and from those sites they can hop further.
This second order virtual process produces an effective hopping of order
$t^2 / 2\varepsilon_0$ between the even sites, so that the effective
Hamiltonian is given by
\begin{equation}
  H_{\rm eff} =
   -{t^2 \over 2\varepsilon_0 }\sum_{j {\rm even},\sigma}
   (c_{j,\sigma}^\dagger c_{j+2,\sigma}^{\vphantom{\dagger}}
	    + {\rm h.c.})
  + U \sum_{j{\rm even}} n_{j,\uparrow}n_{j,\downarrow}
 \;.
\end{equation}
 Here the Hamiltonian acts on a Hilbert space with empty odd sites. This
effective Hamiltonian
describes a half--filled Hubbard model with a hopping amplitude $\tilde t=t^2
/2\varepsilon_0$ and repulsion $\tilde U=U$, so again we can use the expression
of the charge
gap for the Hubbard model. In the weak coupling limit ($U\ll
t^2/\varepsilon_0$), we get
\begin{equation}
    \Delta_c=
	 {8\over \pi} t \sqrt{U \over 2 \varepsilon_0}
	 \exp\left(-\pi t^2 \over U \varepsilon_0\right)
  \label{eq:gapsmall_B}
 \;.
\end{equation}
In the intermediate coupling limit, where $U$ is larger than the effective
hopping,
but smaller then the energy splitting of the two bands $\epsilon_0$, the gap is
given
by $\Delta_c=U - 2t^2 / \varepsilon_0$.

\section{Weak coupling limit}

  In this limit we can first diagonalize the hopping part of the Hamiltonian
and
treat the interaction as a perturbation. To do
  that, we introduce the Fourier transforms of the electron creation and
annihilation operators keeping in mind that there are two sites in the unit
cell:
\begin{eqnarray}
  c_{k,\sigma}^{\vphantom{\dagger}} &=& \sqrt{2\over L}
    \sum_{j {\rm even}} {\rm e}^{-ikj/2} c_{j,\sigma}^{\vphantom{\dagger}}
     \nonumber\\
  \tilde c_{k,\sigma}^{\vphantom{\dagger}} &=& \sqrt{2\over L}
    \sum_{j {\rm odd}} {\rm e}^{-ik(j-1)/2} c_{j,\sigma}^{\vphantom{\dagger}}
  \label{eq:defc_k}
 \;.
\end{eqnarray}
Here $L$ is the number of sites, $L/2$ is the number of unit cells, and the
length of the unit cell is set to 1.

  \subsection{Alternating hopping amplitudes}

 The Fourier transform of the kinetic part of the  Hamiltonian
(\ref{eq:Ham_t_alter}) is
\begin{equation}
   H_0  =
     -\sum_{k,\sigma}
     \left[
       (t_1+t_2 {\rm e}^{-ik})
       c_{k,\sigma}^\dagger
       \tilde c_{k,\sigma}^{\vphantom{\dagger}}
      +(t_1+t_2 {\rm e}^{ik})
	\tilde c_{k,\sigma}^\dagger
	c_{k,\sigma}^{\vphantom{\dagger}}
     \right]
  \label{eq:Ham_t_alter_k}
 \;.
\end{equation}

Let us introduce annihilation operators for the electrons in the lower
($d_{k,\sigma}$) and upper ($f_{k,\sigma}$) bands by
\begin{eqnarray}
  d_{k,\sigma}^{\vphantom{\dagger}}&=&{1\over \sqrt{2}}\left[
    s(k) c_{k,\sigma}^{\vphantom{\dagger}} +
    s^*(k) \tilde c_{k,\sigma}^{\vphantom{\dagger}}
       \right] \;, \nonumber\\
  f_{k,\sigma}^{\vphantom{\dagger}}&=&{1\over \sqrt{2}}\left[
    s(k) c_{k,\sigma}^{\vphantom{\dagger}} -
    s^*(k) \tilde c_{k,\sigma}^{\vphantom{\dagger}}
       \right]
  \label{eq:defdbda}
 \;,
\end{eqnarray}
where $s(k)=\exp{i (\alpha_k/2 +k/4)}$ and
$\alpha_k$ is defined by
\begin{equation}
 \tan\alpha_k = {t_2-t_1 \over t_2+t_1} \tan {k\over 2}
\;,\quad
\left(-{\pi\over 2}<\alpha_k<{\pi\over 2}\right)
 \label{eq:alphak}
 \;.
\end{equation}
The hopping part of the Hamiltonian is now diagonal:
\begin{equation}
  H_0=-\sum_{k,\sigma} \varepsilon(k)
    \left(
      d_{k,\sigma}^\dagger d_{k,\sigma}^{\vphantom{\dagger}} -
      f_{k,\sigma}^\dagger f_{k,\sigma}^{\vphantom{\dagger}}
    \right)
 \;,
\end{equation}
with
\begin{equation}
  \varepsilon(k)=\sqrt{t^2_1+t^2_2 +2 t_1 t_2 \cos k }
 \;.
\end{equation}
 The Fermi velocity $v_F$ is given by
\begin{equation}
  v_F
  ={t_1 t_2 \over \sqrt{t_1^2+t_2^2}}
  ={1\over 8} {W^2-\Delta_D^2 \over \sqrt{(W^2+\Delta_D^2)/2}}
 \;,
\end{equation}
where $\Delta_D= 2(t_1-t_2)$ is the dimerization gap and $W=2(t_1+t_2)$ is the
total bandwidth (see Section.~I).

 The Fourier transform of the Hamiltonian (\ref{eq:Ham_int}) describing the
interaction is
\begin{equation}
    H_{\rm int}  =
       U {1\over L} \sum_{k_1,k_2,k_3,k_4}
      \left(
	 c_{k_1,\uparrow}^\dagger
	 c_{k_2,\downarrow}^\dagger
	 c_{k_3,\downarrow}^{\vphantom{\dagger}}
	 c_{k_4,\uparrow}^{\vphantom{\dagger}}
	 +
	 \tilde c_{k_1,\uparrow}^\dagger
	 \tilde c_{k_2,\downarrow}^\dagger
	 \tilde c_{k_3,\downarrow}^{\vphantom{\dagger}}
	 \tilde c_{k_4,\uparrow}^{\vphantom{\dagger}}
      \right)
  \label{eq:Ham_int_k}
 \;,
\end{equation}
where $k_1+k_2-k_3-k_4 = Q$ is a vector of the reciprocal lattice.
Usually $Q=0$, but if one of the bands is half filled, the umklapp
processes
with $Q=\pm2\pi$ become important.
Using the operators defined by Eq.~(\ref{eq:defdbda}), the interaction can be
written as
\begin{equation}
 H_{\rm int}={U \over 2} {1\over L}
   \sum_{k_1,k_2,k_3,k_4}
	 \cos\left(
	   {\alpha_{k_1}+\alpha_{k_2}-\alpha_{k_3}
	    -\alpha_{k   _4}\over 2}+ {Q \over 4}\right)
	 d_{k_1,\uparrow}^\dagger
	 d_{k_2,\downarrow}^\dagger
	 d_{k_3,\downarrow}^{\vphantom{\dagger}}
	 d_{k_4,\uparrow}^{\vphantom{\dagger}}
   \label{eq:intA0}
 \;,
\end{equation}
where we have kept only the fermions near the Fermi energy.

  From the interaction part of the Hamiltonian we can identify the interactions
between the electrons near the Fermi surface, the so called  $g$-couplings
of the $g$-ology\cite{Solyom}.
 Usually these couplings are spin dependent. However, in our model the
interaction is isotropic, thus we do not have to worry about the spin
dependence,
and they  read:
\begin{eqnarray}
  g_1 &=& g_2 = g_4 = {U\over 2}  \;, \nonumber\\
  g_3 &=& -\sin(2 \alpha_F) {U\over 2} =
{\Delta_D \over W}{2\over 1+(\Delta_D / W)^2} {U\over 2}
   \;.
  \label{eq:geff0}
\end{eqnarray}
The
umklapp scattering amplitude $g_3$ vanishes linearly with $\Delta_D$ for small
$\Delta_D/W$,
in agreement with the
estimates\cite{Seidel} obtained
using perturbational arguments to get the strength
of the umklapp scattering. Clearly, the model of Eq.~(\ref{eq:geff0}) is not
equivalent to the half--filled Hubbard model, for which all the g couplings are
equal and given by $U$.

 In perturbation theory, the logarithmic corrections to the vertex  generate
the differential equations of the renormalization group (RG) approach when one
integrates out the degrees of freedom far from the Fermi level
\cite{Solyom,Emery}.
There are four differential equations altogether, but near half--filling  we
need only the
equations which describe the charge degrees of freedom and give rise to the
charge gap. As we know from Larkin and Sak\cite{larkin}, to account correctly
for the fluctuations which give rise to the $\sqrt{U}$ factor
in Eq.~(\ref{eq:gapt2small}), we
need the RG equations up to third order in $g$-s.
Introducing $g=2g_2-g_1$ and denoting by $\tilde g$ the renormalized couplings
corresponding to momentum cutoff $\tilde D$,  the $3^{\rm rd}$ order
RG equations we need are
\begin{eqnarray}
   { {\rm d} \tilde g \over {\rm d} \ln \tilde D} &=&
     - {1\over \pi v_c} \tilde g_3^2
     +  {1\over 2\pi^2 v_c^2} \tilde g \tilde g_3^2 \;,\nonumber\\
  { {\rm d} \tilde g_3 \over {\rm d} \ln \tilde D} &=&
     -{1\over \pi v_c} \tilde g \tilde g_3
     +  {1\over 4\pi^2 v_c^2}
       \left[ \tilde g^2 \tilde g_3+ \tilde g_3^3 \right]
  \label{eq:scaequ}
 \;,
\end{eqnarray}
where $v_c=v_F+g_4/2\pi$ is the velocity of the charge excitations. There
is a scaling invariant
$  C = (\tilde g^2 - \tilde g_3^2) / (2\pi v_c - \tilde g)$
, and the system of equations above can be rewritten as a single differential
equation:
\begin{eqnarray}
 { {\rm d} \tilde g \over {\rm d} \ln \tilde D} = \pi v_c
   \left(
     {\tilde g^2 \over \pi^2 v_c^2 }
     + {C \tilde g \over \pi^2 v_c^2}  -{2 C\over \pi v_c}
   \right)
  \left( 1- {\tilde g\over2 \pi v_c} \right)
  \label{eq:scaeqred}
 \;.
\end{eqnarray}

  Reducing the cutoff $\tilde D$, we scale towards the strong coupling region,
where the RG equations are no longer valid. This cross--over
occurs when the cutoff $\tilde D$ corresponds to an energy scale that has been
identified with the charge gap\cite{Emery,larkin,Barisic}. So integrating the
scaling
equations (\ref{eq:scaeqred}), we get
\begin{eqnarray}
  \int_{\Delta_c / v_c}^{\tilde D_0} d \ln \tilde D = -
  \int_{\tilde g(\Delta_c)}^g {d \tilde g \over \pi v_c}
   \left(
     {\tilde g^2 \over \pi^2 v_c^2 }
     + {C \tilde g \over \pi^2 v_c^2}  -{2 C\over \pi v_c}
   \right)^{-1}
  \left( 1- {\tilde g\over2 \pi v_c} \right)^{-1}
 \;,
\end{eqnarray}
where $\tilde D_0$ is the initial cutoff with $\tilde g(\tilde D_0)=g$ and
$\tilde g_3(\tilde D_0)=g_3$. Keeping only the leading and next to leading
terms of $g$, and  introducing the notation $\xi=g_3/g$, one gets
\begin{equation}
  \ln{\Delta_c\over v_c}-\ln \tilde D_0= {1\over2}\ln {\xi g\over\pi v_c}-
    \left({ \pi v_c \over g} - {1\over 4} \right)
    {{\rm Arth}\sqrt{1-\xi^2}\over \sqrt{1-\xi^2}} +
    {1\over 4} + \ln C_\Delta
 \;,
\end{equation}
where the constant $\ln C_\Delta$ contains the terms like
$1/\tilde g(\Delta_c)$ which are small compared to $1/g$, as $\tilde
g$ scales towards the strong coupling limit. As in that region the
RG equations are not any more valid, all the possible corrections are
incorporated in that constant
\cite{larkin}. We have separated $1/4$ from the constant $\ln C_\Delta$ for
convenience.

Replacing $v_c$ by $v_F+g/2\pi$ ( $g_4=g$ in the leading order in $U$), and
exponentiating, we get
\begin{equation}
  \Delta_c  = C_\Delta \tilde D_0 v_F \sqrt{\xi g\over\pi v_F}
    \exp\left(-{ \pi v_F \over g}
      {{\rm Arth}\sqrt{1-\xi^2}\over \sqrt{1-\xi^2}}
    \right)
    \exp\left( - {1\over 4}
    {{\rm Arth}\sqrt{1-\xi^2}\over \sqrt{1-\xi^2}}+{1\over 4}
    \right)
   \label{eq:gapsca}
 \;.
\end{equation}
Its behavior in the case of small dimerization is given by
\begin{equation}
    \Delta_c= C_\Delta \tilde D_0 v_F
      \sqrt{2 g \over \pi v_F }
      e^{1/4}
     \left( \xi \over 2 \right)^{
     \left(
	{ \pi v_F /g } + {3 / 4}
     \right)}
  \label{eq:gaplim}
 \;,
\end{equation}
where we have used the fact that ${\rm Arth} \sqrt{1-\xi^2} \approx \ln(2/\xi)$
if
$\xi \rightarrow 0$. This is the same form as found by Luther\cite{Luther76}
for the charge gap if the umklapp scattering is small compared to the
$2g_2-g_1$, which in our notation means $\xi\ll 1$ (see also
Ref.~[\onlinecite{Emery}]).

Using the g-couplings of Eq.~(\ref{eq:geff0}) in Eq.~(\ref{eq:gapsca}) provides
us with
the functional dependence of the gap on the repulsion $U$. This expression also
includes
a dependence on $t_2/t_1$. It does {\it not} give the full dependence on
$t_2/t_1$ however.
 The reason is
the following:
Our estimate of the charge gap accounts neither for the curvature of
the band (as the RG equations are based on a model with
linear dispersion relation) nor for the possible processes involving the
empty band. This virtual processes will contribute as higher order
corrections in $U$ to the effective interactions near
 the Fermi surface. Since in the formula for the charge
gap these effective couplings appear essentially as $\exp(-\pi v_F /g)$, the
$U^2$ correction to the effective coupling will give an $O(1)$ correction, i.e.
if we assume for a moment that $g=U-\gamma U^2 + O(U^3)$, then
$1/g=1/U+\gamma+O(U)$ and $\exp(-\pi v_F /g)=\exp(-\pi v_F \gamma)\exp(-\pi v_F
/U)$. So the correction appears as a multiplying factor that depends on
$t_2/t_1$ in the equation for the gap. Clearly, to get a quantitative estimate
of the gap one has to include these contributions.

To be fully consistent with our determination of the $U$ dependence, which
relies on
RG equations up to third order in g (see Eq.~(\ref{eq:scaequ})), one should
in principle
calculate the vertex corrections to the coupling constants to third order in
$U$. The problem with that program is that the calculation of the third order
is hopelessly cumbersome. We think however that a calculation of the vertex
correction to second order in $U$ is sufficient.
The procedure we have used is the following: First, we calculate the
second order corrections to the
effective interactions near the Fermi surface by
integrating out all the electron states except those which are closer to the
Fermi surface than some
small cutoff $\tilde D_0$ in the lower band. Then,
if the cut-off  $\tilde D_0$ is small enough,
 the
dispersion relation of the
electrons within this cut-off around the Fermi surface is essentially
linear and we can safely use the
RG equations of Eq.~(\ref{eq:scaequ}) to
decrease the cutoff further down to $\Delta_c / v_c$.
Now, the cut-off  $\tilde D_0$ can be taken very small: It just has to be
larger
than $\Delta_c / v_c$, and we already know that this quantity is exponentially
small in the weak coupling limit. So this procedure should be valid. Finally,
we will
see that for
small $\tilde D_0$ ($W/ v_c \gg \tilde D_0 \gg \Delta_c / v_c$) the result for
the gap is
independent of the cutoff $\tilde D_0$.

To calculate the higher corrections in $U$, we must take into account the
virtual
processes that involve states in the
upper band and consider the full interaction Hamiltonian instead of
Eq.~(\ref{eq:intA0}):
\begin{eqnarray}
 H_{\rm int}&=&{U \over 2} {1\over L}
   \sum_{k_1,k_2,k_3,k_4}
	 \cos\left(
	   {\alpha_{k_1}+\alpha_{k_2}-\alpha_{k_3}
	    -\alpha_{k   _4})\over 2}+ {Q \over 4}\right) \nonumber\\
      &\times &\left(
	 d_{k_1,\uparrow}^\dagger
	 d_{k_2,\downarrow}^\dagger
	 d_{k_3,\downarrow}^{\vphantom{\dagger}}
	 d_{k_4,\uparrow}^{\vphantom{\dagger}}
       + f_{k_1,\uparrow}^\dagger
	 f_{k_2,\downarrow}^\dagger
	 d_{k_3,\downarrow}^{\vphantom{\dagger}}
	 d_{k_4,\uparrow}^{\vphantom{\dagger}}
       + d_{k_1,\uparrow}^\dagger
	 d_{k_2,\downarrow}^\dagger
	 f_{k_3,\downarrow}^{\vphantom{\dagger}}
	 f_{k_4,\uparrow}^{\vphantom{\dagger}} + \dots
     \right)
   \label{eq:intA}
 \;,
\end{eqnarray}
where we have written only the terms which are responsible for the
second order corrections in $U$.
To that order, the effective interactions are given by
\begin{eqnarray}
  g_1(\tilde D_0)&=&
    {U\over 2} - {U^2 \over 4} \left(D_1 + D_3 \right) \;,
  \nonumber\\
  g_2(\tilde D_0)&=&
    {U\over 2} - {U^2 \over 4} \left(D_1 + D_2 + D_3 \right) \;,
  \nonumber\\
  g_3(\tilde D_0)&=&
    -\sin(2 \alpha_F) {U\over 2} - {U^2 \over 4} \left(D_4 + D_5 \right)
 \;,
\end{eqnarray}
where we have denoted by $D_1,...,D_5$ the diagrams contributing to the
effective interaction (see Fig.~\ref{fig:perturbA}). They are given by
\begin{eqnarray}
  D_1 &=&
   2  \int_{\pi/2+\tilde D_0}^{\pi} {dq \over 2 \pi}
   {1 \over 2 \varepsilon_F - 2 \varepsilon(q)}
  +2  \int_{0}^{\pi/2-\tilde D_0} {dq \over 2 \pi}
   {1 \over 2 \varepsilon(q) - 2 \varepsilon_F}
  \;,\nonumber\\
  D_2 &=&
   - 2  \int_{0}^{\pi/2-\tilde D_0} {dq\over 2 \pi}
   {\cos^2(\alpha_F-(\alpha_q+\alpha_{\pi-q})/2)
   +\sin^2(\alpha_F+(\alpha_q+\alpha_{\pi-q})/2)
   \over
   \varepsilon(q) -  \varepsilon(\pi-q) }
  \;,\nonumber\\
  D_3 &=&
   2  \int_{0}^{\pi} {dq\over 2 \pi}
   {1 \over 2 \varepsilon_F + 2 \varepsilon(q)}
  \;,\nonumber\\
  D_4 &=&
   4  \int_{0}^{\pi/2-\tilde D_0} {dq\over 2 \pi}
   {-\cos(\alpha_F-(\alpha_q+\alpha_{\pi-q})/2)
    \sin(\alpha_F+(\alpha_q+\alpha_{\pi-q})/2)
   \over
  2\varepsilon_F+ \varepsilon(q) +  \varepsilon(\pi-q) }
  \;,\nonumber\\
  D_5 &=&
   -4  \int_{0}^{\pi/2} {dq \over 2 \pi}
   {-\cos(\alpha_F-(\alpha_q+\alpha_{\pi-q})/2)
    \sin(\alpha_F+(\alpha_q+\alpha_{\pi-q})/2)
   \over
   \varepsilon(q) - \varepsilon(\pi-q) }
  \;.
\end{eqnarray}
Here $\alpha_F$ and $\varepsilon_F$ stands for $\alpha(k_F)$ and
$\varepsilon(k_F)$.

The integrals $D_1$, $D_2$ and $D_5$ contains a logarithmic singularity
$\ln(2 / \tilde D_0)$, and, after some algebra, one can separate these
contributions, so that
the effective interactions can be written as
\begin{eqnarray}
  g_1(\tilde D_0) &=&
    {U\over 2} - {U^2 \over 4}{1 \over \pi v_F}
    \left[
      \ln{ 2 \over \tilde D_0} + {A\over 2} I_0^+(A) +O(\tilde D_0)
    \right] \;,
  \nonumber\\
  g_2(\tilde D_0) &=&
    {U\over 2} - {U^2 \over 4}{1 \over \pi v_F}
    \left\{
     {1-\sin^2( 2\alpha_F) \over 2}
     \ln{ 2 \over  \tilde D_0} +
     {A \over 4} \left[ I_0^+(A) + I_0^-(A) \right] \right.
     \nonumber\\
     &&\left.-\sin^2(2 \alpha_F) {A \over 4}
      \left[ I_0^+(A) - I_0^-(A) \right]
     - {A\over 4} \sin^2( 2\alpha_F) I_1(A) +O(\tilde D_0)
    \right\} \;,
  \nonumber\\
  g_3(\tilde D_0) &=&-\sin(2 \alpha_F)
    \left(
      {U\over 2} - {U^2 \over 4}{1 \over \pi v_F}
      \left\{
       -\ln{ 2 \over  \tilde D_0}
       -{A \over 2} \left[ I_0^+(A) - I_0^-(A) \right]
      \right.
     \right.
     \nonumber\\
     &-&
     \left.
     \left.
     {A\over 4}  I_1(A) + {A\over 2}  I_2(A) +O(\tilde D_0)\right\}
    \right) \;,
   \label{eq:gIs}
\end{eqnarray}
where $A=2 t_1 t_2 / (t_1^2 +t_2^2)$ and $I_0^\pm$, $I_1$ and $I_2$ are
non--singular integrals given by:
\begin{eqnarray}
 I_0^\pm(A) &=&
  \int_0^{\pi/2} d\varphi
    {1 \over 1 + \sqrt{1\pm A \cos(\varphi)} }
  \;,\nonumber\\
 I_1(A) &=&
  \int_0^{\pi/2} d\varphi
    \left(
     {1 \over \sqrt{1- A \cos(\varphi)}}
    -{1 \over \sqrt{1+ A \cos(\varphi)}}
    \right)
  \;,\nonumber\\
 I_2(A) &=&
  \int_0^{\pi/2} d\varphi
    \left(1+
     {1 \over \sqrt{1- A^2 \cos^2(\varphi)}}
    \right)
    {1 \over 2
      + \sqrt{1+ A \cos(\varphi)}
      +\sqrt{1- A \cos(\varphi)}} \;.\nonumber\\
\end{eqnarray}
  For $A \ll 1$, they read:
\begin{eqnarray}
 I_0^\pm(A) &=& {\pi \over 4} \mp {A \over 8} + O(A^2)
   \;,\nonumber\\
 I_1(A) &=&  A + O(A^3)
  \;,\nonumber\\
 I_2(A) &=&  \pi/4 + O(A^2)
  \;.
\end{eqnarray}
 If it were only for a single band with linear dispersion relation these
integrals near the  $\ln(2 / \tilde D_0)$ would not appear.


  Using the couplings of Eq.~(\ref{eq:gIs}) in Eq.~(\ref{eq:gapsca}), we get
\begin{eqnarray}
  \Delta_c  &=& 2 C_\Delta  v_F
    \sqrt{2 U  \over \pi v_F } (1-A^2)^{1/4}
    \exp\left(-{ 2\pi v_F \over U}
      {{\rm Arth} A\over A}
    \right)
    \nonumber\\
   &&\times
    \exp\left( - {1\over 4}
    {{\rm Arth}A\over A}+{1\over 4}
    \right)
    e^{\tilde C(A)} [1+O(\tilde D_0)]
  \label{eq:gap_sUAnd}
 \;,
\end{eqnarray}
where, to first order,  $\tilde D_0$  disappeared from the expression of the
gap.
This expression now describes
the full functional dependence of $\Delta_c$ on $U/t_1$ and $t_2/t_1$. The
only thing that remains is
to determine the prefactor $C_\Delta$. This can be done as follows: From the
Bethe ansatz
solution of the Hubbard model,
 we know the value of
the gap exactly in the limit of large dimerization
[see Eq.~(\ref{eq:gapt2small})]. It is easily checked that in the limit $t_2
\rightarrow 0$ the functional dependence of Eq.~(\ref{eq:gap_sUAnd}) is the
same as in
Eq.~(\ref{eq:gapt2small}). So we can use the result of
Eq.~(\ref{eq:gapt2small}) to determine the prefactor, and this gives
$C_\Delta=2 \sqrt{2/\pi}$. Our final expression for the gap in the weak
coupling limit is thus
\begin{eqnarray}
  \Delta_c  &=& { 4 v_F \over \pi} \sqrt{ U  \over v_F } (1-A^2)^{1/4}
    \exp\left(-{ 2\pi v_F \over U}
      {{\rm Arth} A\over A}
    \right)
    \nonumber\\
   &&\times
    \exp\left( - {1\over 4}
    {{\rm Arth}A\over A}+{1\over 4}
    \right)
    e^{\tilde C(A)}
  \label{eq:gap_sUArd}
 \;,
\end{eqnarray}
where $A=2t_1t_2/(t_1^2+t_2^2)$ and $\tilde C(A)$ is given by
\begin{eqnarray}
  \tilde C(A) &=&
    {1\over 4A}
   \left[ 2 (1-A^2)I^-_0(A) + 2 A^2 I_0^+(A) - (1-2 A^2) I_1(A)
	 -2 I_2(A) \right] \nonumber \\
   &+&
    {{\rm Arth} A \over 4A^2}
    \left[ -2 I^-_0(A) + (1- A^2) I_1(A)
	 + 2 (1- A^2) I_2(A) \right]
  \label{eq:gap_csUA}
 \;.
\end{eqnarray}

It is interesting to compare this approach with that of Larkin and
Sak\cite{larkin}, who used slightly different arguments when they calculated
the gap in the negative $U$ Hubbard model in the small $U$ limit. It can be
shown very easily that our method is equivalent to theirs and gives the same
result for the attractive Hubbard model.

\subsection{Alternating on--site energies}

 The Fourier transform of the kinetic Hamiltonian (\ref{eq:Ham_e_alter}) using
the Fourier transforms of the operators defined in Eq.~(\ref{eq:defc_k}) is
\begin{equation}
   H_0  =
     -\varepsilon_0 \sum_{k,\sigma}
     \left(
       c_{k,\sigma}^\dagger
       c_{k,\sigma}^{\vphantom{\dagger}}
      - \tilde c_{k,\sigma}^\dagger
	\tilde c_{k,\sigma}^{\vphantom{\dagger}}
     \right)
    - 2 t \cos(k/2) \sum_{k,\sigma}
     \left(
       {\rm e}^{-ik/2}
       c_{k,\sigma}^\dagger
       \tilde c_{k,\sigma}^{\vphantom{\dagger}}
      + {\rm e}^{ik/2}
	\tilde c_{k,\sigma}^\dagger
	c_{k,\sigma}^{\vphantom{\dagger}}
     \right)
  \label{eq:Ham_e_alter_k}
  \;.
\end{equation}
It can be diagonalized in terms of  new creation and annihilation
operators defined by
\begin{eqnarray}
  d_{k,\sigma}^{\vphantom{\dagger}}&=&
    c_{k,\sigma}^{\vphantom{\dagger}}\cos \beta_k +
      {\rm e}^{-ik/2}
    \tilde c_{k,\sigma}^{\vphantom{\dagger}} \sin \beta_k
    \;,\nonumber\\
  f_{k,\sigma}^{\vphantom{\dagger}}&=&
    c_{k,\sigma}^{\vphantom{\dagger}}\sin \beta_k -
    {\rm e}^{-ik/2}
    \tilde c_{k,\sigma}^{\vphantom{\dagger}}\cos \beta_k
  \label{eq:defdldu}
 \;,
\end{eqnarray}
where $\beta_k$ is given by
\begin{equation}
 \tan 2\beta_k = {2 t\over \varepsilon_0}\cos {k\over 2} \;,\quad
\left({-\pi\over 4}<\beta_k<{\pi\over 4}\right)
 \label{eq:betak}
 \;.
\end{equation}
The kinetic part of the Hamiltonian reads
\begin{equation}
  H_0=-\sum_{k,\sigma} \varepsilon(k)
    \left(
      d_{k,\sigma}^\dagger
      d_{k,\sigma}^{\vphantom{\dagger}} -
      f_{k,\sigma}^\dagger
      f_{k,\sigma}^{\vphantom{\dagger}}
    \right)
 \;,
\end{equation}
with
\begin{equation}
  \varepsilon(k)=\sqrt{\varepsilon_0^2+ 2t^2 +2 t^2 \cos k }
 \;.
\end{equation}
 The dimerization gap opening at the Brillouin zone
boundary is $\Delta_D= 2\varepsilon_0$,
 the 'total' bandwidth is
$W=2 \sqrt{\varepsilon_0^2+4 t^2}$ and the Fermi velocity  is
\begin{equation}
  v_F= {t^2 \over \sqrt{\varepsilon_0^2+2 t^2}}
 \;.
\end{equation}

 Replacing the new operators defined in Eq.~(\ref{eq:defdldu}) for the lower
and upper band into the interaction Hamiltonian
(\ref{eq:Ham_int_k}) we get:
\begin{eqnarray}
  H_{\rm int}&=&U  {1\over L}
   \sum_{k_1,k_2,k_3,k_4} \left[ \right.
 \nonumber\\
     && \left(
     \cos \beta_{k_1}
     \cos \beta_{k_2}
     \cos \beta_{k_3}
     \cos \beta_{k_4}
    \pm
     \sin \beta_{k_1}
     \sin \beta_{k_2}
     \sin \beta_{k_3}
     \sin \beta_{k_4}
     \right)
	 d_{k_1,\uparrow}^\dagger
	 d_{k_2,\downarrow}^\dagger
	 d_{k_3,\downarrow}^{\vphantom{\dagger}}
	 d_{k_4,\uparrow}^{\vphantom{\dagger}}
 \nonumber\\
     &+& \left(
     \sin \beta_{k_1}
     \cos \beta_{k_2}
     \cos \beta_{k_3}
     \cos \beta_{k_4}
    \mp
     \cos \beta_{k_1}
     \sin \beta_{k_2}
     \sin \beta_{k_3}
     \sin \beta_{k_4}
     \right)
     \nonumber \\
    && \times \left(
	 f_{k_1,\uparrow}^\dagger
	 d_{k_2,\downarrow}^\dagger
	 d_{k_3,\downarrow}^{\vphantom{\dagger}}
	 d_{k_4,\uparrow}^{\vphantom{\dagger}}
       +
	 f_{k_1,\downarrow}^\dagger
	 d_{k_2,\uparrow}^\dagger
	 d_{k_3,\uparrow}^{\vphantom{\dagger}}
	 d_{k_4,\downarrow}^{\vphantom{\dagger}}
       +
	 d_{k_4,\uparrow}^\dagger
	 d_{k_3,\downarrow}^\dagger
	 d_{k_2,\downarrow}^{\vphantom{\dagger}}
	 f_{k_1,\uparrow}^{\vphantom{\dagger}}
       +
	 d_{k_4,\downarrow}^\dagger
	 d_{k_3,\uparrow}^\dagger
	 d_{k_2,\uparrow}^{\vphantom{\dagger}}
	 f_{k_1,\downarrow}^{\vphantom{\dagger}}
      \right) \nonumber\\
     &+& \left(
     \cos \beta_{k_1}
     \cos \beta_{k_2}
     \sin \beta_{k_3}
     \sin \beta_{k_4}
    \pm
     \cos \beta_{k_1}
     \cos \beta_{k_2}
     \sin \beta_{k_3}
     \sin \beta_{k_4}
     \right)
     \nonumber \\
     &&\times
     \left(
	 f_{k_1,\uparrow}^\dagger
	 f_{k_2,\downarrow}^\dagger
	 d_{k_3,\downarrow}^{\vphantom{\dagger}}
	 d_{k_4,\uparrow}^{\vphantom{\dagger}}
    +
	 d_{k_4,\uparrow}^\dagger
	 d_{k_3,\downarrow}^\dagger
	 f_{k_2,\downarrow}^{\vphantom{\dagger}}
	 f_{k_1,\uparrow}^{\vphantom{\dagger}}
\right)
 \nonumber\\
     &+& \dots \left. \right]
 \;,
\end{eqnarray}
with the constraint $k_1+k_2-k_3-k_4=Q$ and the upper (lower) signs stand for
the normal $Q=0$ (umklapp, $Q=\pm 2\pi$) scattering, respectively. There are
altogether 16 terms, but we have kept only the terms which we need to calculate
the $g$'s near the Fermi surface up to second order in $U$.

  Like in the case of alternating hopping amplitudes, to determine the
effective couplings $g$, we first integrate out the high energy processes
up to some cutoff $\tilde D_0$:\begin{eqnarray}
  g_1(\tilde D_0) &=&
    U (\cos^4 \beta_F + \sin^4 \beta_F )
     - U^2 \left(D_1 + 2 D_2 + D_3 +2 D_4 \right) \;,
  \nonumber\\
  g_2(\tilde D_0) &=&
    U (\cos^4 \beta_F + \sin^4 \beta_F )
    - U^2  \left(D_1 + 2 D_2 + D_3 +D_5 + 2D_6 \right) \;,
  \nonumber\\
  g_3(\tilde D_0) &=&
    U (\cos^4 \beta_F - \sin^4 \beta_F )
  - U^2 \left(2 D_7 + D_8 + D_9 + 2 D_{10} \right)
 \;,
\end{eqnarray}
where we have denoted by $D_1,...,D_{10}$ the diagrams contributing to the
effective interactions (see Fig.~\ref{fig:perturbB}) which are given by
\begin{eqnarray}
  D_1 &=&
   2  \int_{\pi/2+\tilde D_0}^{\pi} {dq  \over 2 \pi}
   { \left(
      \cos^2 \beta_F \cos^2 \beta_q + \sin^2 \beta_F \sin^2 \beta_q
     \right)^2
   \over 2 \varepsilon_F - 2 \varepsilon(q)}
   \nonumber\\
  &+& 2  \int_{0}^{\pi/2-\tilde D_0} {dq  \over 2 \pi}
   {\left(
      \cos^2 \beta_F \cos^2 \beta_q + \sin^2 \beta_F \sin^2 \beta_q
    \right)^2
    \over 2 \varepsilon(q) - 2 \varepsilon_F}
  \;,\nonumber\\
  D_2 &=&
    2  \int_{\pi/2}^{\pi} {dq \over 2 \pi}
   {\left(
      \cos^2 \beta_F \cos \beta_q \sin \beta_q
    - \sin^2 \beta_F \sin \beta_q \cos \beta_q
     \right)^2
    \over
  2 \varepsilon_F }
  \;,\nonumber\\
  D_3 &=&
   2  \int_{0}^{\pi} {dq \over 2 \pi}
   {\left(
      \cos^2 \beta_F \sin^2 \beta_q + \sin^2 \beta_F \cos^2 \beta_q
    \right)^2
   \over 2 \varepsilon_F + 2 \varepsilon(q)}
  \;,\nonumber\\
  D_4 &=&
   -2  \int_{0}^{\pi/2} {dq \over 2 \pi}
   {\left(
      \cos^2 \beta_F \sin \beta_q \cos \beta_q
    - \sin^2 \beta_F \sin \beta_q \cos \beta_q
     \right)^2
   \over
  2 \varepsilon(q) }
  \;,\nonumber\\
  D_5 &=&
   -2  \int_{0}^{\pi/2-\tilde D_0} {dq  \over 2 \pi}
   {  2 \cos^4 \beta_F \cos^2 \beta_q \cos^2 \beta_{\pi-q}
     + 2 \sin^4 \beta_F \sin^2 \beta_q \sin^2 \beta_{\pi-q}
   \over  \varepsilon(q) -  \varepsilon(\pi-q) }
  \;,\nonumber\\
  D_6 &=&
     -\int_{0}^{\pi/2} {dq  \over 2 \pi}
   {  2 \cos^4 \beta_F \cos^2 \beta_q \sin^2 \beta_{\pi-q}
    + 2 \sin^4 \beta_F \sin^2 \beta_q \cos^2 \beta_{\pi-q}
   \over \varepsilon(q) +  \varepsilon(\pi-q) }
  \;,\nonumber\\
  D_7 &=&
   2  \int_{\pi/2}^{\pi} {dq \over 2 \pi}
   { \cos^4 \beta_F \cos^2 \beta_q \sin^2 \beta_{\pi-q}
    - \sin^4 \beta_F \sin^2 \beta_q \cos^2 \beta_{\pi-q}
   \over
  2 \varepsilon_F -\varepsilon(q) + \varepsilon(\pi-q) }
  \;,\nonumber\\
  D_8 &=&
   2  \int_{0}^{\pi} {dq \over 2 \pi}
   {  \cos^4 \beta_F \sin^2 \beta_q \sin^2 \beta_{\pi-q}
    - \sin^4 \beta_F \cos^2 \beta_q \cos^2 \beta_{\pi-q}
   \over
  2 \varepsilon_F +\varepsilon(q) + \varepsilon(\pi-q) }
  \;,\nonumber\\
  D_9 &=&
   -4  \int_{0}^{\pi/2-\tilde D_0} {dq \over 2 \pi}
   {  \cos^4 \beta_F \cos^2 \beta_q \cos^2 \beta_{\pi-q}
    - \sin^4 \beta_F \sin^2 \beta_q \sin^2 \beta_{\pi-q}
   \over
   \varepsilon(q) - \varepsilon(\pi-q) }
  \;,\nonumber\\
  D_{10} &=&
   -2  \int_{0}^{\pi/2} {dq \over 2 \pi}
   {  \cos^4 \beta_F \cos^2 \beta_q \sin^2 \beta_{\pi-q}
    - \sin^4 \beta_F \sin^2 \beta_q \cos^2 \beta_{\pi-q}
   \over
   \varepsilon(q) + \varepsilon(\pi-q) }
  \;.
\end{eqnarray}

 Again, like for the case of alternating hoppings, we separate the logarithmic
divergencies from integrals $D_1$, $D_5$ and $D_9$, and we write the effective
couplings as
\begin{eqnarray}
   g_1(\tilde D_0)&=& {\varepsilon_0^2 +t^2 \over \varepsilon_0^2 +2 t^2} U
       - {U^2 \over \pi v_F}
       \left[
	 \left(
	   \varepsilon_0^2 +t^2 \over \varepsilon_0^2 +2 t^2
	 \right)^2
     \ln {2\over \tilde D_0} + I_1 \right] \;, \nonumber\\
   g_2(\tilde D_0)&=& {\varepsilon_0^2 +t^2 \over \varepsilon_0^2 +2 t^2}U
       - {U^2 \over \pi v_F}
      \left( {1\over 2} {t^4 \over (\varepsilon_0^2 + 2t^2)^2}
	 \ln {2\over \tilde D_0} +\tilde I_2
      \right) \;,\nonumber\\
   g_3(\tilde D_0)&=& {\varepsilon_0  \over \sqrt{\varepsilon_0^2 +2 t^2}}
       U
       - {U^2 \over \pi v_F}
     \left[-{\varepsilon_0 (\varepsilon_0^2 +t^2)
	 \over (\varepsilon_0^2 +2 t^2)^{3/2}}
      \ln {2\over \tilde D_0} + \tilde I_3 \right]
 \;.
\end{eqnarray}
The expressions for the $\tilde I$'s are rather complicated and, for brevity,
we
will give here only their value in the limit of large dimerization
(the compounds which can be described by this model presumably
have parameters which
lie in this limit):
\begin{eqnarray}
    \tilde I_1 &=& {32+13 \pi \over 32} {t^6 \over \varepsilon_0^6}
		 + O(t^8/\varepsilon_0^8) \;, \nonumber\\
    \tilde I_2 &=& {t^4 \over 4 \varepsilon_0^4} -
	  {48+3 \pi \over 32} {t^6 \over \varepsilon_0^6}
		 + O(t^8/\varepsilon_0^8) \;, \nonumber\\
    \tilde I_3 &=& {t^2 \over 4  \varepsilon_0^4} -
	  {48+5 \pi \over 32} {t^6 \over \varepsilon_0^6}
		+ O(t^8/\varepsilon_0^8)
   \label{eq:IsB}
 \;.
\end{eqnarray}
We have calculated numerically the $\tilde I$-s for a few values of
$\varepsilon_0/t$ and they are presented in Table~\ref{tab:Ival}.

For $g_4$, it is sufficient to take the value without the $U^2$ corrections:
\begin{equation}
  g_4 = U {\varepsilon_0^2 +t^2 \over \varepsilon_0^2 +2 t^2}
 \;.
\end{equation}

Replacing the effective interactions into Eq.~(\ref{eq:gapsca}), we get
\begin{eqnarray}
  \Delta_c  &=&
    {8 \over \pi} t
   \sqrt{U \varepsilon_0 \over 2(\varepsilon_0^2 + 2 t^2)}
   \exp\left(-{\pi  \over U} \sqrt{\varepsilon_0^2 + 2 t^2}
      {\rm Arth} {t^2 \over \varepsilon_0^2 + t^2}\right)
   \exp\left[
     {1 \over 4}-{t^2 +\varepsilon_0^2 \over 4t^2}
	{\rm Arth} { t^2 \over \varepsilon_0^2 + t^2}
   \right]
   e^{ \tilde C}
   \label{eq:gap_smallUB}
 \;,
  \nonumber\\
\end{eqnarray}
where the constant has been fixed so that for large dimerization
($\varepsilon_0 \gg t$) we get Eq.~(\ref{eq:gapsmall_B}). Here $\tilde C$ is
given by
\begin{eqnarray}
  \tilde C &=& (2 \tilde I_2 - \tilde I_1)
   {\left( 2 t^2 + \varepsilon_0^2 \right)^2 \over t^4}
   \left[
      1-{ t^2 +\varepsilon_0^2 \over t^2}
      {\rm Arth} { t^2 \over \varepsilon_0^2 + t^2}
   \right] \nonumber \\
   &-& \tilde I_3
  { ( 2 t^2 + \varepsilon_0^2)^{3/2}
   \over
    t^2\varepsilon_0 }
     \left[
      {\varepsilon_0^2 + t^2 \over t^2}-
     {\varepsilon_0^2(2 t^2 +\varepsilon_0^2) \over t^4}
      {\rm Arth} { t^2 \over \varepsilon_0^2 + t^2}
   \right]
 \label{eq:defC}
 \;.
\end{eqnarray}

\section{Large U limit}

  In the large $U$ limit a canonical transformation can be applied to obtain an
effective Hamiltonian analog to the $t-J$ Hamiltonian of the non dimerized
Hubbard model. We know already that
if there are two alternating hoppings, a dimerization gap $\Delta_D$ opens
at the Brillouin zone boundary, and, in the limit $U\rightarrow +\infty$,
this gap becomes the charge gap. This will be
modified by the spin interaction, which will give corrections of order
$1/U$.

\subsection{Alternating hoppings}

 If $U\gg t_1,t_2$, and since we have two different hopping amplitudes, the
effective
model will be a $t-J$ model with alternating $t_1$ and $t_2$ hoppings and
alternating $J_1$ and $J_2$ exchange interactions:
\begin{eqnarray}
H_{tJ} &=& -
  t_1 \sum_{i {\rm even}, \sigma}
    \tilde {\cal P}
    (c_{i,\sigma}^\dagger   c_{i+1,\sigma}^{\vphantom{\dagger}} + {\rm h.c.})
    \tilde {\cal P}
    -t_2 \sum_{i {\rm odd}, \sigma}
    \tilde {\cal P}
    (c_{i,\sigma}^\dagger c_{i+1,\sigma}^{\vphantom{\dagger}} + {\rm h.c.})
    \tilde {\cal P}
    \nonumber\\
  &+&
    J_1 \sum_{i {\rm even}}
      (S_{i} S_{i+1}-{1\over 4} n_i n_{i+1})
    +J_2 \sum_{i {\rm odd}}
	(S_{i} S_{i+1}-{1\over 4} n_i n_{i+1})
     \nonumber\\
&+&
  {J\over 4} \sum_{i,\sigma}
  \tilde {\cal P}
  (c_{i,\sigma}^\dagger c_{i+1,-\sigma}^\dagger
   c_{i+1,\sigma}^{\vphantom{\dagger}}  c_{i+2,-\sigma}^{\vphantom{\dagger}}
   -
   c_{i,\sigma}^\dagger c_{i+1,-\sigma}^\dagger
   c_{i+1,-\sigma}^{\vphantom{\dagger}} c_{i+2,\sigma}^{\vphantom{\dagger}}
   + {\rm h.c.})
   \tilde {\cal P}
   + {\cal O}(t^3/U^2)
 \;.
\end{eqnarray}
Here the projector
$\tilde {\cal P}=\prod_i[1-n_{i,\uparrow}n_{i,\downarrow}] $
 ensures that there are no doubly occupied sites,
$n_i=n_{i,\uparrow}+n_{i,\downarrow}$, and the generated exchange couplings are
$J_1=4t_1^2/U$, $J_2=4t_2^2/U$ and $J=4 t_1 t_2/U$.

  To go further, let us follow the scheme developed by Shiba and Ogata in their
study of the correlation functions for the large $U$ Hubbard
model\cite{ShibaOgata}.
The basic idea is to treat the exchange part as a perturbation. This is
possible if the
exchange integrals are smaller than the hopping integrals, which reduces to the
condition
$J_1 \ll t_2$ (remember that $t_2<t_1$, and thus that $J,J_2<J_1$), or, in
terms
of the original parameters, $U\gg 4t_1^2/t_2$.
In this limit,
 the part describing essentially hopping of spinless fermions can be solved
exactly and the exchange part is then treated as a perturbation. This can be
achieved by assuming that the wave function is the
product of a charge and a spin wave function, where the spin wave function is
defined in a Hilbert space of dimension $2^N$, i.e every charge has the
additional freedom of having its spin up or down:
\begin{equation}
  |\Psi \rangle = |sf \rangle \otimes |\Phi\rangle
 \;.
\end{equation}
Here $|\Phi\rangle$ is the spin part of the wave function, while $|sf \rangle
$,
the wave function of the N spinless fermions, is the ground state of the
kinetic Hamiltonian. Following standard perturbation technique for the case of
a degenerate ground state, we will have to diagonalize the following $2^N\times
2^N$ Hamiltonian to lift the degeneracy of  $|\Phi\rangle$ and get the $1/U$
energy corrections:
\begin{eqnarray}
\langle H_{tJ} \rangle' / L
  = &-& t_1 \langle c_{0}^\dagger c_{1}^{\vphantom{\dagger}} \rangle'
    - t_2 \langle c_{1}^\dagger c_{2}^{\vphantom{\dagger}} \rangle'
    \nonumber \\
   &+& \left(
      {J_1 \over 2} \langle n_0 n_1 \rangle'
    + {J_2 \over 2} \langle n_1 n_2 \rangle'
    - J \langle
      c_0^\dagger c_1^\dagger
      c_1^{\vphantom{\dagger}}  c_2^{\vphantom{\dagger}}
     \rangle'
   \right)
   {1 \over N} \sum_{j=1}^N (S_j S_{j+1}-{1\over 4})
  \label{eq:e_tJ}
 \;,
\end{eqnarray}
where the $\langle \hat A \rangle'$ denotes the expectation value of the
operator $\hat A$ in the Fermi sea of spinless fermions, i.e. $\langle \hat A
\rangle'=\langle sf| \hat A |sf \rangle$.

  Before we continue with the calculation of the gap, we need to know the
expectation values of electron operators in the Fermi see
of spinless fermions, where $k_{F,{\rm sf}}=2
k_F$. Using the translational symmetry of the system, it is enough to calculate
the
following basic expectation values
\begin{eqnarray}
  \langle c_{0}^\dagger c_{j}^{\vphantom{\dagger}} \rangle' &=&
    {2\over L}\sum_k {\rm e}^{ikj/2}
    \langle c_k^\dagger c_k^{\vphantom{\dagger}} \rangle'
   \;,\nonumber\\
  \langle c_{1}^\dagger c_{j+1}^{\vphantom{\dagger}} \rangle' &=&
    {2\over L}\sum_k {\rm e}^{ikj/2}
    \langle \tilde c_k^\dagger \tilde c_k^{\vphantom{\dagger}} \rangle'
  \;,\nonumber\\
  \langle c_{1}^\dagger c_{j}^{\vphantom{\dagger}} \rangle' &=&
    {2\over L}\sum_k {\rm e}^{ikj/2}
    \langle \tilde c_k^\dagger c_k^{\vphantom{\dagger}} \rangle'
  \label{eq:cocjsumk}
 \;,
\end{eqnarray}
where $j$ is even everywhere.  More complicated expressions can
be calculated by using Wick's theorem, i.e. by constructing all
possible pairings. For example:
\begin{eqnarray}
  \langle n_1 n_j \rangle' &=&
  \langle c_{1}^\dagger c_{1}^{\vphantom{\dagger}} \rangle'
  \langle c_{j}^\dagger c_{j}^{\vphantom{\dagger}} \rangle'
  -\langle c_{1}^\dagger c_{j}^{\vphantom{\dagger}} \rangle'
  \langle c_{j}^\dagger c_{1}^{\vphantom{\dagger}} \rangle'
  = n^2 - \langle c_{1}^\dagger c_{j}^{\vphantom{\dagger}} \rangle'^2
  \;,\nonumber\\
  \langle
      c_0^\dagger c_1^\dagger
      c_1^{\vphantom{\dagger}}  c_2^{\vphantom{\dagger}}
  \rangle'&=&
  n \langle c_0^\dagger c_2^{\vphantom{\dagger}} \rangle'
  -\langle c_0^\dagger c_1^{\vphantom{\dagger}} \rangle'
  \langle c_1^\dagger c_2^{\vphantom{\dagger}} \rangle'
   \;.
  \label{eq:wick}
\end{eqnarray}

 In the following, the sums  above will
be replaced by integrals,
\begin{equation}
  {2\over L}\sum_k \rightarrow {1 \over 2\pi} \int_{-\pi}^{\pi} dk
 \;.
\end{equation}

The free fermion Hamiltonian~(\ref{eq:Ham_t_alter_k}) has been solved in
Sec.~VI. Using the diagonal
electron operators $d$ and $f$ defined in
Eq.~(\ref{eq:defdbda}), we get
\begin{eqnarray}
     \langle c_k^\dagger c_k^{\vphantom{\dagger}} \rangle' &=&
     \langle \tilde c_k^\dagger \tilde c_k^{\vphantom{\dagger}} \rangle'
     = {1\over 2}(n_{d,k} + n_{f,k}) \nonumber \\
    \langle \tilde c_k^\dagger c_k^{\vphantom{\dagger}} \rangle' &=&
     {s^{*2}(k)\over 2}(n_{d,k} - n_{f,k})
 \;,
\end{eqnarray}
where $n_{d,k}=d_{k}^\dagger d_{k}^{\phantom{\dagger}}$, similarly for
$n_{f,k}$, and the spin index of the fermion operators has been dropped. For
$j$ even, this gives
\begin{eqnarray}
  \langle c_{0}^\dagger c_{j}^{\vphantom{\dagger}} \rangle' =
      \langle c_{1}^\dagger c_{j+1}^{\vphantom{\dagger}} \rangle' &=&
    {1 \over 2\pi} \int_{-\pi}^{\pi} dk
    {\rm e}^{ikj/2}{1\over 2}(n_{b,k} + n_{f,k})  \nonumber\\
  &=& {\sin  \pi j n \over \pi j}
  \label{eq:cocj}
 \;.
\end{eqnarray}
This is the same expression as for a non--dimerized model. For
$\langle c_{1}^\dagger c_{j}^{\vphantom{\dagger}} \rangle'$, we get:
\begin{equation}
\langle c_{1}^\dagger c_{j}^{\vphantom{\dagger}} \rangle'
= {1\over 2\pi}\int_0^{2\varphi_0} dk \cos(k(j-1)/2-\alpha_k)
 \;,
\end{equation}
where $\alpha_k$ is defined in Eq.~(\ref{eq:alphak}) and
\begin{equation}
    \varphi_0 =
    \left\{
      \begin{array}{ll}
	 \pi n & \mbox{, if $n < 1/2$;} \\
	 \pi (1-n)& \mbox{, if $n > 1/2$.}
      \end{array}
    \right.
      \label{eq:varphi0}
\end{equation}
After some algebra, we get
\begin{equation}
\langle c_{1}^\dagger c_{j}^{\vphantom{\dagger}} \rangle'
= {1\over \pi}\int_0^{\varphi_0} d\varphi
  {t_1 \cos \varphi j  + t_2 \cos \varphi(j-2)
   \over
   \sqrt{t^2_1+t^2_2 +2 t_1 t_2 \cos 2\varphi}}
 \;.
\end{equation}
For $j=0$ and $j=2$, this can be written
\begin{eqnarray}
  \langle c_{1}^\dagger c_{0}^{\vphantom{\dagger}} \rangle'
  &=& {1\over 2\pi t_1}
  \left[
    (t_1-t_2) F(\varphi_0,q)+(t_1+t_2) E(\varphi_0,q)
  \right]
   \;, \nonumber\\
  \langle c_{1}^\dagger c_{2}^{\vphantom{\dagger}} \rangle'
  &=& {1\over 2\pi t_2}
  \left[
    (t_2-t_1) F(\varphi_0,q)+(t_1+t_2) E(\varphi_0,q)
  \right]
  \label{eq:expval01A}
 \;,
\end{eqnarray}
where we have introduced the notation
\begin{equation}
  q={2\sqrt{t_1 t_2}\over t_1+t_2}
  \label{eq:defqA}
 \;,
\end{equation}
and where the elliptic integrals $E(\varphi_0,q)$ and $F(\varphi_0,q)$ are
defined by
\begin{eqnarray}
   F(\varphi_0,q)&=&\int_0^{\varphi_0}
     {d \varphi \over \sqrt{1-q^2 \sin\varphi^2} }
   \;, \nonumber\\
   E(\varphi_0,q)&=&\int_0^{\varphi_0}
     {d \varphi \sqrt{1-q^2 \sin\varphi^2} }
  \label{eq:ellint}
 \;.
\end{eqnarray}

Now, let us turn back to the effective Hamiltonian (\ref{eq:e_tJ}).
For the spin degrees of freedom, we are left with an $N$--site Heisenberg
Hamiltonian with an effective exchange interaction $J_{\rm eff}$ which we can
read from Eq.~(\ref{eq:e_tJ}):
\begin{eqnarray}
J_{\rm eff} &=&
  {J_1 \over 2} \langle n_0 n_1 \rangle'
    + {J_2 \over 2} \langle n_1 n_2 \rangle'
    - J \langle
      c_0^\dagger c_1^\dagger
      c_1^{\vphantom{\dagger}}  c_2^{\vphantom{\dagger}}
     \rangle' \nonumber\\
 &=& {2\over U}
 \left[
   n^2 (t_1^2+t_2^2)-n t_1 t_2 {\sin 2\pi n \over \pi}
   -{(t_1-t_2)^2 \over \pi^2} F^2(\varphi_0,q)
  \right]
 \;,
\end{eqnarray}
where we have used Eqs.~(\ref{eq:wick}) and (\ref{eq:expval01A})

 In the ground state of the AF Heisenberg model the expectation value $\langle
\Phi |(S_0 S_1-{1\over 4})| \Phi \rangle$ is $-\ln 2$ in zero magnetic field,
as shown by Griffiths \cite{griffiths}.
The energy per site $\langle \Psi | H_{tJ} | \Psi \rangle / L$ is then
\begin{equation}
\varepsilon_{tJ}
  = \varepsilon_{sf} - J_{\rm eff} \ln 2
 \label{eq:e_tJ2}
 \;,
\end{equation}
where
$\varepsilon_{\rm sf}=
-t_1 \langle c_{1}^\dagger c_{0}^{\vphantom{\dagger}} \rangle'
-t_2 \langle c_{1}^\dagger c_{2}^{\vphantom{\dagger}} \rangle'$
is the kinetic energy of the spinless fermions per site. From
Eq.~(\ref{eq:expval01A}), we get
\begin{equation}
 \varepsilon_{\rm sf}(n)=-{t_1+t_2\over \pi} E(\varphi_0,q)
 \label{eq:ensf}
 \;,
\end{equation}
which in the non--dimerized case
($q=1$) gives the correct energy $(2t/\pi) \sin{\pi n}$.

For $n=1/2$, we get
\begin{equation}
 J_{\rm eff} = {2\over U}
 \left[ {t_1^2+t_2^2 \over 4}-{(t_1-t_2)^2 \over \pi^2} K^2(q)
  \right]
 \;,
\end{equation}
where $K(q)=F(\pi/2,q)$ is the complete elliptic integral defined in
Eq.~(\ref{eq:ellint}).

  The energy $E_{tJ}/L$ has a cusp as a function of filling at $n=1/2$, and
this nonanaliticity comes from the $\varphi_0$ [see Eq.~(\ref{eq:varphi0})].
Therefore the first derivative of $E_{tJ}/L$ has a jump at quarter filling,
which is nothing but the charge gap:
\begin{equation}
  \Delta_c =
      \left. \partial \varepsilon_{tJ} \over \partial n \right\vert_{n=1/2+0}
     -\left. \partial \varepsilon_{tJ} \over \partial n \right\vert_{n=1/2-0}
  \label{eq:gapcont}
 \;.
\end{equation}
In our case this gives
\begin{equation}
  \Delta_c = 2(t_1-t_2) \left[1- {4 \ln 2 \over \pi}
    {t_1+t_2 \over U} K\left({2\sqrt{t_1 t_2}\over t_1+t_2}\right)\right]
  \label{eq:gap_lUAnd}
 \;,
\end{equation}
where we have used the following identities:
\begin{eqnarray}
  \left. \partial F(\varphi_0,q) \over \partial n \right\vert_{n=1/2+0}
  -\left. \partial F(\varphi_0,q) \over \partial n \right\vert_{n=1/2-0}
  &=& -{2 \pi \over \sqrt{1-q^2}}
  \;, \nonumber\\
  \left. \partial E(\varphi_0,q) \over \partial n \right\vert_{n=1/2+0}
  -\left. \partial E(\varphi_0,q) \over \partial n \right\vert_{n=1/2-0}
  &=& -{2 \pi \sqrt{1-q^2} }
 \;.
\end{eqnarray}

Using the limiting behavior of the elliptic function
$K(q)=\ln(4/\sqrt{1-q^2})$, the gap in the small dimerization limit is
\begin{equation}
  \Delta_c = 2(t_1-t_2) \left[1- {4 \ln 2 \over \pi}
    {t_1+t_2 \over U} \ln{4(t_1+t_2)\over t_1-t_2}\right]
 \;,
\end{equation}
or
\begin{equation}
  \Delta_c = \Delta_D \left[1- {2 \ln 2 \over \pi}
    {W \over U} \ln{4 W \over \Delta_D}\right]
  \label{eq:gap_lUsdim}
 \;.
\end{equation}

  In Fig.~\ref{fig:gaplargeu} we have compared the analytical results presented
above with the numerical data. The agreement is already very good at
$U/t_1=10$ for
$t_2/t_1$ between 0.3 and 0.5.  To get such a good agreemnt
for $t_2/t_1=0.1$, one has to go to larger values
of $U$, so that the condition $U\gg 4 t_1^2/t_2 \approx 40$ is satisfied.
For large values of $t_2$, the gap is small
and the numerical estimate is not accurate.

  As $t_2 \rightarrow 0$, Eq.~(\ref{eq:gap_lUAnd}) gives
 $\Delta_c= 2 t_1-(4 \ln 2) t_1^2 /U$, in apparent contradiction with
the second of equations
(\ref{eq:gapt2is0}), which gives $\Delta_c= 2 t_1 -4 t_1^2  /U$ for $t_2=0$.
The origin of the discrepancy is that Eq.~(\ref{eq:gap_lUAnd}) holds if
 $U \gg t_1^2/t_2$, while the second of equations (\ref{eq:gapt2is0})
requires that   $t_2 \ll t_1^2/U$.
These conditions are clearly incompatible.
 Actually the  cross--over between these two regimes
can be observed on Fig.~\ref{fig:gaplargeu}, where the
points below the solid line
($t_1^2/U\approx 0.1$) gives $\Delta_c\sim 2 t_1-4 t_1^2 /U$, while for
larger values of $U$ we get the correct $4\ln 2$ slope.


\subsection{Alternating on--site energies}

  Like in the previous case, a canonical transformation can be applied in the
limit $U\gg t,\epsilon_0$ to obtain a $t-J$ like
effective Hamiltonian
\begin{eqnarray}
H_{tJ} &=& -
  t \sum_{i , \sigma}
    \tilde {\cal P}(c_{i,\sigma}^\dagger   c_{i+1,\sigma}^{\vphantom{\dagger}}
+ {\rm h.c.})
\tilde {\cal P}
   -\varepsilon_0 \sum_{i {\rm even}, \sigma}
     (c_{i,\sigma}^\dagger
      c_{i,\sigma}^{\vphantom{\dagger}}
      - c_{i+1,\sigma}^\dagger
       c_{i+1,\sigma}^{\vphantom{\dagger}})
  \nonumber\\
  &+&
  J \sum_{i }
    (S_{i} S_{i+1}-{1\over 4} n_i n_{i+1})
    \nonumber\\
&+&
  {J\over 4} \sum_{i,\sigma}
  \tilde {\cal P}
  (c_{i,\sigma}^\dagger c_{i+1,-\sigma}^\dagger
   c_{i+1,\sigma}^{\vphantom{\dagger}}  c_{i+2,-\sigma}^{\vphantom{\dagger}}
   -
   c_{i,\sigma}^\dagger c_{i+1,-\sigma}^\dagger
   c_{i+1,-\sigma'}^{\vphantom{\dagger}} c_{i+2,\sigma}^{\vphantom{\dagger}}
   + {\rm h.c.})
   \tilde {\cal P}
   + {\cal O}(t^3/U^2)  \;,
\end{eqnarray}
where the projector
$\tilde {\cal P}=\prod_i[1-n_{i,\uparrow}n_{i,\downarrow}] $
 ensures that there are no doubly occupied sites and the generated exchange
coupling is
$J=4t^2/U$.
If the additional condition $J\ll \epsilon_0$ is satisfied (that is, if
$U\gg 4t^2/\epsilon_0$), we can again follow Ogata and Shiba.

Let us now calculate the expectation values for this model. The free fermion
model is given by Eq.~(\ref{eq:Ham_e_alter_k}) and it
has been solved
 using transformations Eq.~(\ref{eq:defdldu}). So, for the expectation
values we get
\begin{eqnarray}
  \langle c_k^\dagger c_k^{\vphantom{\dagger}} \rangle' &=&
    {n_{d,k} + n_{f,k} \over 2}
    + {n_{d,k} - n_{f,k} \over 2} \cos 2\beta_k
  \;,\nonumber \\
  \langle
    \tilde c_k^\dagger \tilde c_k^{\vphantom{\dagger}}
  \rangle' &=&
    {n_{d,k} + n_{f,k} \over 2}
    - {n_{d,k} - n_{f,k} \over 2} \cos 2\beta_k
  \;,\nonumber \\
    \langle \tilde c_k^\dagger c_k^{\vphantom{\dagger}} \rangle' &=& {\rm
e}^{-ik/2}
   {n_{d,k} - n_{f,k} \over 2} \sin 2\beta_k
  \nonumber \\
    \langle c_k^\dagger \tilde c_k^{\vphantom{\dagger}} \rangle' &=& {\rm
e}^{ik/2}
   {n_{d,k} - n_{f,k} \over 2} \sin 2\beta_k
 \;.
\end{eqnarray}
Using the results we obtained in Eq.~(\ref{eq:cocj}), for $j$ even we get
\begin{eqnarray}
  \langle c_{0}^\dagger c_{j}^{\vphantom{\dagger}} \rangle' &=&
  {\sin  \pi j n \over \pi j} + I_j
  \nonumber\\
  \langle c_{1}^\dagger c_{j+1}^{\vphantom{\dagger}} \rangle'
  &=& {\sin  \pi j n \over \pi j} - I_j
  \;,
\end{eqnarray}
and after some algebra
\begin{equation}
 \langle c_{0}^\dagger c_{j-1}^{\vphantom{\dagger}} \rangle' =\langle
c_{1}^\dagger c_{j}^{\vphantom{\dagger}} \rangle' =
  {t\over \varepsilon_0 } (I_j + I_{j-2})
  \label{eq:expval01B}
 \;,
\end{equation}
where the $I_j$ denote the following integral:
\begin{eqnarray}
I_j &=&{1\over 2} {1\over 2\pi}
   \int_{2 \varphi_0}^{2 \varphi_0} dk
   {\rm e}^{ikj/2} \cos 2\beta_k    \nonumber\\
   &=& {1\over 2\pi}\int_0^{2 \varphi_0 } dk
  { \cos(kj/2) \over \sqrt{1+ (2t/\varepsilon_0)^2 \cos^2(k /2) } }
 \;,
\end{eqnarray}
and  $\varphi_0$ is defined by Eq.~(\ref{eq:varphi0}). These are elliptic
integrals and e.g.  $I_0$ and $I_2$ are given by
\begin{eqnarray}
  I_0 &=& {1\over \pi} {\varepsilon_0 \over \sqrt{\varepsilon_0^2+4 t^2}}
    F(\varphi_0,q)
  \;, \nonumber\\
  I_0-I_2 &=& {2\over \pi}
  {\varepsilon_0  \sqrt{\varepsilon_0^2+4 t^2} \over 4 t^2}
   \left[ F(\varphi_0,q)- E(\varphi_0,q)\right]
  \label{eq:I0I2}
 \;,
\end{eqnarray}
where
\begin{equation}
q={2t\over\sqrt{\varepsilon_0^2+4t^2}}
 \;.
\end{equation}

The energy of the system is $E_{tJ}=\langle \Psi | H_{tJ} | \Psi \rangle$ and
the energy per site $\varepsilon_{tJ}=E_{tJ}/ L$  is still given by
Eq.~(\ref{eq:e_tJ2}), but now the kinetic energy of the spinless fermions per
site is
\begin{eqnarray}
 \varepsilon_{\rm sf}(n) &=&
  2 t
   \langle
     c_{0}^\dagger c_{1}^{\vphantom{\dagger}}
   \rangle'
   - {1\over 2} \left(
      \langle
	c_{0}^\dagger c_{0}^{\vphantom{\dagger}}
      \rangle'
     -\langle
	c_{1}^\dagger c_{1}^{\vphantom{\dagger}}
      \rangle'
   \right)
    \nonumber \\
 &=&- {\sqrt{\varepsilon_0^2+4 t_2} \over \pi} E(\varphi_0,q)
 \;,
\end{eqnarray}
where we have used Eqs.~(\ref{eq:expval01B}) and (\ref{eq:I0I2}).  For the
effective exchange
coupling we get
\begin{eqnarray}
J_{\rm eff} &=&
  J \left[
       \langle n_0 n_1 \rangle'
     -{1\over 2}
     \left(
     \langle
      c_0^\dagger c_1^\dagger
      c_1^{\vphantom{\dagger}}  c_2^{\vphantom{\dagger}}
     \rangle'
    + \langle
      c_1^\dagger c_2^\dagger
      c_2^{\vphantom{\dagger}}  c_3^{\vphantom{\dagger}}
     \rangle'
    \right)
   \right] \nonumber \\
  &=& {4 t^2 \over U}
   \left\{
     n^2-n {\sin 2 \pi n \over 2 \pi}
    - {\varepsilon_0^2 \over 2 \pi^2 t^2}
      F(\varphi_0,q) \left[ F(\varphi_0,q)-E(\varphi_0,q) \right]
   \right\}
 \;,
\end{eqnarray}
where we have used Wick's theorem, namely:
\begin{eqnarray}
  \langle n_0 n_1 \rangle' &=& n_0 n_1
  -\langle c_{0}^\dagger c_{1}^{\vphantom{\dagger}} \rangle'^2
\;,\nonumber\\
 \langle
      c_0^\dagger c_1^\dagger
      c_1^{\vphantom{\dagger}}  c_2^{\vphantom{\dagger}}
 \rangle'&=&
  n_1 \langle c_0^\dagger c_2^{\vphantom{\dagger}} \rangle'
  -\langle c_0^\dagger c_1^{\vphantom{\dagger}} \rangle'^2
   \;,\nonumber\\
 \langle
      c_1^\dagger c_2^\dagger
      c_2^{\vphantom{\dagger}}  c_3^{\vphantom{\dagger}}
 \rangle'&=&
  n_0 \langle c_1^\dagger c_3^{\vphantom{\dagger}} \rangle'
  -\langle c_0^\dagger c_1^{\vphantom{\dagger}} \rangle'^2
   \;,
\end{eqnarray}
and
\begin{eqnarray}
  \langle n_0 n_1 \rangle' - {1\over 2}
  \left(
  \langle
      c_0^\dagger c_1^\dagger
      c_1^{\vphantom{\dagger}}  c_2^{\vphantom{\dagger}}
 \rangle'
  +\langle
      c_1^\dagger c_2^\dagger
      c_2^{\vphantom{\dagger}}  c_3^{\vphantom{\dagger}}
 \rangle'
 \right) &=&
  n_0 n_1
  - {1\over 2} ( n_1 \langle c_0^\dagger c_2^{\vphantom{\dagger}} \rangle'
   + n_0 \langle c_1^\dagger c_3^{\vphantom{\dagger}} \rangle')
  \nonumber\\
  &=& n^2-n {\sin 2\pi n \over 2 \pi} -I_0(I_0-I_2)
   \;.
\end{eqnarray}

For $n=1/2$, we get
\begin{equation}
 J_{\rm eff} = {1  \over U}
   \left\{
     t^2
    - {2 \varepsilon_0^2 \over  \pi^2 }
      K(q) \left[ K(q)-E(\pi/2,q) \right]
   \right\}
 \;.
\end{equation}

  The gap, using Eq.~(\ref{eq:gapcont}), is
\begin{equation}
  \Delta_c = 2 \varepsilon_0 \left\{1- {2 \ln 2 \over \pi}
    {\sqrt{\varepsilon_0^2+4 t^2} \over U}
  \left[(1+q^2) K(q)-E(\pi/2,q)\right]
  \right\}
  \label{eq:gaplargeUB}
 \;,
\end{equation}
and in the small dimerization limit it is given by
\begin{equation}
  \Delta_c = 2 \varepsilon_0 \left[1- {4 \ln 2 \over \pi}
    {\sqrt{\varepsilon_0^2+4 t^2} \over U}
  \ln { 4 \sqrt{\varepsilon_0^2+4 t^2} \over \varepsilon_0 }
  \right]
 \;.
\end{equation}
If we parametrize the gap in this limit with $\Delta_D$ and $W$, we can see
that it is identical with Eq.~(\ref{eq:gap_lUsdim}).

\acknowledgements
  We would like to thank D. Baeriswyl,
H. Beck, E. Jekelman, T. M. Rice, J. S\'olyom and A.
Zawadowski for stimulating discussions.
  One of the author (K.P.) acknowledges the financial support of the
Swiss National Foundation No 20-33964.92 and 2000-037653.93/1

\begin{figure}
\caption{ Schematic representation of the model with: a) alternating hopping
amplitudes $t_1$ and $t_2$; b) alternating on--site energies with
energy splitting $2\varepsilon_0$ .
\label{fig:model}}
\end{figure}

\begin{figure}
\caption{ Band structure of the model with $U=0$.
\label{fig:bands}}
\end{figure}

\begin{figure}
\caption{ Schematic picture of the different regions as a function of model
parameters for the model with: a) alternating hopping amplitudes; b)
alternating
on--site energies.
\label{fig:region}}
\end{figure}

\begin{figure}
\caption{ a) Approximation for the charge gap for $t_2/t_1=$0.1 to 0.9 in
increments of 0.1 from the top to the bottom. The diamonds are the numerical
data for $t_2/t_1=$0.1 to 0.5;
b) Approximation for the gap, but now only from $t_2/t_1=$0.6 to 0.9 in
increments of 0.1 from the top to the bottom.
  \label{fig:appr}}
\end{figure}

\begin{figure}
\caption{
  Numerical estimates of the gap (diamonds) for $t_2=0.1$ compared with the
analytical results.
\label{fig:largedim}}
\end{figure}

\begin{figure}
\caption{ Weak coupling expression compared to numerical estimates for
$t_2=0.3$ (upper curve) and $t_2=0.4$ (lower curve).
  \label{fig:gapweak}}
\end{figure}

\begin{figure}
\caption{
  Numerical values of the gap $\Delta_c(L)$ for L=4,8,12 and 16 as a
function of $1/L$ for $U/t_1=5$. Note that
$\Delta_c(L)$ converges very well for small values of
$t_2/t_1$.
\label{fig:gapsca}}
\end{figure}

\begin{figure}
\caption{ Perturbational corrections to the effective backward scattering
$g_1$, forward scattering $g_2$ and umklapp scattering $g_3$ for
the model with alternating hoppings shown up to the second order. The solid
(dashed) line represents the fermions in the lower $d$ (upper $f$) band.
\label{fig:perturbA}}
\end{figure}

\begin{figure}
\caption{ Perturbational corrections to the effective backward scattering
$g_1$, forward scattering $g_2$ and umklapp scattering $g_3$ for
the model with alternating hoppings shown up to the second order. The solid
(dashed) line represents the fermions in the lower $d$ (upper $f$) band.
\label{fig:perturbB}}
\end{figure}

\begin{figure}
\caption{ Numerical estimates of the gap in the large $U$
limit (diamonds) together with the analytical approximation. The results
presented are for $t_2=$0.1, 0.2, 0.3, 0.4, 0.5 and 0.7 from the top to bottom
   \label{fig:gaplargeu}}
\end{figure}

\begin{table}
\caption{ The parameters of the gap equations for the weak coupling and strong
coupling limit for the model with alternating hoppings.\label{tab:althop}}
\begin{tabular}{ddddd}
$t_2/t_1$ & $a$ & $b$ & $\Delta_D/t_1$ & $c$ \\
\hline
0.1 & 0.36887 &  0.6336 & 1.8 & 1.6816 \\
0.2 & 0.46782 &  1.2990 & 1.6 & 2.0167 \\
0.3 & 0.49990 &  2.0304 & 1.4 & 2.3984 \\
0.4 & 0.48604 &  2.8669 & 1.2 & 2.8368 \\
0.5 & 0.43672 &  3.8588 & 1.0 & 3.3474 \\
0.6 & 0.36100 &  5.0790 & 0.8 & 3.9555 \\
0.7 & 0.26867 &  6.6518 & 0.6 & 4.7075 \\
0.8 & 0.17046 &  8.8399 & 0.4 & 5.7054 \\
0.9 & 0.07709 & 12.4449 & 0.2 & 7.2658 \\
\end{tabular}
\end{table}

\begin{table}
\caption{ The parameters $U_0$, $a_1$, $b_1$ and $b_2$.
\label{tab:pade}}
\begin{tabular}{dddddd}
$t_2/t_1$ & $U_0/t_1$ & $a_1$ & $b_1$ & $b_2$ \\
\hline
0.2 & 1.30 &  0.45312 &  1.5438 &  13.0786\\
0.3 & 2.03 & -0.13375 &  1.6173 &  13.5738\\
0.4 & 2.87 & -0.97220 &  1.5540 &  10.7001\\
0.5 & 3.86 & -1.91825 &  1.4287 &   3.3094\\
0.6 & 4.   & -1.84743 &  2.6357 &   0.0736\\
0.7 & 4.   & -1.50580 &  5.3353 &  -2.2707\\
0.8 & 4.   & -1.09045 & 11.5364 &  -7.4996\\
0.9 & 4.   & -0.60222 & 33.3189 & -34.0969\\
\end{tabular}
\end{table}

\begin{table}
\caption{ The parameters of the gap equations for the weak coupling and strong
coupling limit for the model with alternating on--site energies.
\label{tab:altene}}
\begin{tabular}{ddddd}
$\varepsilon_0/t$ & $a$ & $b$ & $\Delta_D/t$ & $c$ \\
\hline
0.2 & 0.60292 & 8.82123 & 0.4 & 5.63087\\
0.4 & 0.78553 & 6.00854 & 0.8 & 4.41422\\
0.6 & 0.90155 & 4.53739 & 1.2 & 3.71014\\
0.8 & 0.97686 & 3.61670 & 1.6 & 3.21882\\
1. & 1.02292 & 2.98899 & 2. & 2.84615\\
2. & 1.03186 & 1.56009 & 4. & 1.78537\\
3. & 0.93780 & 1.04544 & 6. & 1.28117\\
4. & 0.84788 & 0.78494 & 8. & 0.99167\\
5. & 0.77446 & 0.62816 & 10. & 0.80616\\
\end{tabular}
\end{table}

\begin{table}
\caption{ The  $\tilde I$-s for some values of $\varepsilon_0/t$.
\label{tab:Ival}}
\begin{tabular}{dddd}
$\varepsilon_0/t$ & $\tilde I_1$ & $\tilde I_2$ & $\tilde I_3$  \\
\hline
0   & $(1/8)\ln 2$ & 0 & 0 \\
0.2 & 0.0832216 & 0.0007691 & 0.0020951 \\
0.4 & 0.0739705 & 0.0026394 & 0.0039468 \\
0.6 & 0.0613605 & 0.0046787 & 0.0053306 \\
0.8 & 0.0480877 & 0.0061374 & 0.0061143 \\
1   & 0.0360938 & 0.0067624 & 0.0063123 \\
2   & 0.0066400 & 0.0038914 & 0.0035338 \\
3   & 0.0013135 & 0.0015298 & 0.0014380 \\
4   & 0.0003309 & 0.0006442 & 0.0006189 \\
5   & 0.0001034 & 0.0003044 & 0.0002961 \\
\end{tabular}
\end{table}
\end{document}